\documentclass[ reprint, amsmath,amssymb, nofootinbib, aps]{revtex4-1}

\usepackage{graphicx}
\usepackage{dcolumn}
\usepackage{bm}
\usepackage{hyperref}

\usepackage{longtable}
\usepackage[caption=false]{subfig}
\usepackage{array}
\usepackage{color}
\usepackage{colortbl}
\usepackage{multirow}

\renewcommand{\arraystretch}{1.2}

\providecommand{\U}[1]{\protect\rule{.1in}{.1in}}

\definecolor{Cor1}{gray}{0.95}
\newcolumntype{L}[1]{>{\raggedright\let\newline\\\arraybackslash\hspace{0pt}}m{#1}}
\newcolumntype{C}[1]{>{\centering\let\newline\\\arraybackslash\hspace{0pt}}m{#1}}
\newcolumntype{R}[1]{>{\raggedleft\let\newline\\\arraybackslash\hspace{0pt}}m{#1}}
\ifx\pdfoutput\relax\let\pdfoutput=\undefined\fi
\newcount\msipdfoutput
\ifx\pdfoutput\undefined\else
\ifcase\pdfoutput\else
\ifx\paperwidth\undefined\else
\ifdim\paperheight=0pt\relax\else\pdfpageheight\paperheight\fi
\ifdim\paperwidth=0pt\relax\else\pdfpagewidth\paperwidth\fi
\fi\fi\fi

\begin{document}

\title{Observational constraints to a unified cosmological model}

\author{R. R. Cuzinatto}
 \altaffiliation{E-mail: rodrigo.cuzinatto@unifal-mg.edu.br}

\author{E. M. de Morais}%
 \email{E-mail: eduardomessiasdemorais@hotmail.com}
\affiliation{%
Instituto de Ci\^{e}ncia e Tecnologia, Universidade Federal de Alfenas, Rodovia Jos\'{e} Aur\'{e}lio Vilela, 11999, Cidade Universit\'{a}ria, CEP 37715-400, Po\c{c}os de Caldas, MG, Brazil
}%


\author{L. G. Medeiros}
 \homepage{E-mail: leogmedeiros@ect.ufrn.br}
\affiliation{%
 Escola de Ci\^{e}ncia e Tecnologia, Universidade Federal do Rio Grande do Norte, Campus Universit\'{a}rio, s/n, CEP 59072-970, Natal, Brazil
}%

\date{\today}

\begin{abstract}
We propose a phenomenological unified model for dark matter and dark energy based on an equation of state parameter $w$ that scales with the $\arctan$ of the redshift. The free parameters of the model are three constants: $\Omega_{b0}$, $\alpha$ and $\beta$. Parameter $\alpha$ dictates the transition rate between the matter dominated era and the accelerated expansion period. The ratio $\beta / \alpha$ gives the redshift of the equivalence between both regimes. Cosmological parameters are fixed by observational data from Primordial Nucleosynthesis (PN), Supernovae of the type Ia (SNIa), Gamma-Ray Bursts (GRB) and Baryon Acoustic Oscillations (BAO). The calibration of the 138 GRB events is performed using the 580 SNIa of the Union2.1 data set and a new set of 79 high-redshift GRB is obtained. The various sets of data are used in different combinations to constraint the parameters through statistical analysis. The unified model is compared to the $\Lambda$CDM model and their differences are emphasized.
\begin{description}
\item[PACS numbers]
95.35.+d, 95.36.+x, 98.80.Es

\end{description}
\end{abstract}

\maketitle

\section{\label{sec:level1}Introduction}

The recent technological improvement in the space observations deeply altered
Cosmology. In the end of the 90's, the measurement of the luminosity distance of
type Ia supernovae (SNIa) unveiled an accelerating cosmic expansion at recent
times \cite{Riess1998,Perlmutter1999}. This result was later confirmed by other
works using different sets of data \cite{AccelaratingU} such as the cosmic
microwave background radiation (CMB) \cite{Planck2013}, baryon acoustic
oscillations (BAO) \cite{Percival2009,Anderson2014} and even the
relatively recent Gamma-Ray burst (GRB) data \cite{Wei2,Schaefer,Amati1}. As
long as one assumes a homogeneous and isotropic cosmological background, the
cosmic acceleration at low redshifts seems an indisputable observational truth.\footnote{Inhomogeneous cosmological model, such as those in
Refs. \cite{Ellis2012,Tolman1934,Bondi1947,Buchert2000,Rasanen2011,Wiltshire2007,Wiltshire2013}, present alternative explanation to the
apparent present-day cosmic acceleration.}

The simplest theoretical way of describing cosmic acceleration is through the
cosmological constant $\Lambda$, a negative energy density uniformly
distributed throughout the cosmos. The resulting $\Lambda$CDM cosmological
model \cite{Weinberg2008}\ is robust when confronted to observational data,
although it is not a comfortable solution mainly due to the lack of a clear
interpretation of the physical meaning of $\Lambda$ in terms of the known
fundamental interactions. This very fact relegates $\Lambda$ to the mysterious
\textquotedblleft dark sector\textquotedblright\ of the universe. It is
completed by the gravitationally bound cold dark matter (CDM), whose nature is
also unknown.

In face of these two unexplained components, one is tempted to unify them in a
single dark fluid. This unified model would have to be capable of
accelerating the universe at recent times and also provide a dust dominated
epoch toward the past in order to accommodate structure formation. This is our
motivation to introduce the Unified Model (UM) described in Sect. \ref{Sec-UM}.
There is a plethora of cosmological models based on the same idea
\cite{Bertacca2010a}; they are build either based on theoretical motivations
\cite{Peebles1999,Kamenshchik2001,Bento2002,Padmanabhan2002,Scherrer2004,Giannakis2005,Bertacca2007,Giannantonio2006,Santos2006,Radicella2014,Sharif2012} or on phenomenological ones \cite{Makler2003A,Makler2003B,Alcaniz2003,Balbi2007,Pietrobon2008,Ribamar2007,Ribamar2012,Velten2011,Wang2013}. Our
model is built on phenomenological grounds.

The UM is a dynamical model developed from a specific functional form chosen
for the parameter $w$\ of the equation of state $p=w\rho$, where $p$\ is the
pressure related to the cosmic component of density $\rho$: $w$ is given in terms of the
$\arctan$ function. This way, the universe filled with the unified fluid
passes smoothly from a matter-like behavior $(w\simeq0)$ to a dark-energy-like
dynamics $(w\simeq-1)$. This property is justified theoretically once the
history of the universe demands a matter-dominated era with decelerated
expansion ($2\lesssim z\lesssim10000$) followed by an accelerated period
dominated by dark energy ($z\lesssim2$) \cite{Ellis2012}. Our goal is to treat
dark matter and dark energy on the same footing.

The unified scenario for the dark components is meaningful only if one can
constraint the free parameters of the Unified Model by using a large number of
observational data. For this end, we will use the already mentioned SNIa, BAO
and GRB data plus information on the baryon density parameter $\Omega_{b0}%
$\ coming from primordial nucleosynthesis (PN) data.\footnote{More on
$\Omega_{b0}$\ and PN bellow (Sect. \ref{Sec-PN}).}

We used Union2.1 compilation \cite{Union21}\ for obtaining
the distance modulus $\mu$ of the supernovae as a function of their redshift
$z$. For the GRB, we employed data in Ref. \cite{Liu2014b}, which include
29\ GRB in addition to the set of 109 GRB of Ref. \cite{Wei3}. Also, we payed
special attention to the construction of the calibration curve of the GRB. The
procedure involved an interpolation to the points in the plot of $\mu$ as
function of $z$. We noticed that the common interpolation methods, such as
linear and cubic interpolation techniques, are not the best-quality ones. In
fact, Akima's method \cite{Akima} is the one which provides a curve that
naturally connects the observational points without bumps or discontinuities.
We devoted special care to the GRB data as they rise as new
good candidates for standard candles at very high redshifts, with great
potential of revealing additional cosmological information.

The paper is organized as follows. Sect. \ref{Sec-SetUp} presents our Unified
Model (UM) for the dark sector of the universe; in addition, the basic equation of
the $\Lambda$CDM model are reviewed. This prepares the ground for data fitting
aiming to constraint the free parameters of both UM and $\Lambda$CDM. The statistical treatment is performed in Sect. \ref{Sec-Data} after the cosmological data sets used in our analysis have been discussed. The physical consequences of the data fit for the various combinations of data
(PN, SNIa, GRB and BAO) are also addressed in Sect. \ref{Sec-Data} and
further discussed in Sect. \ref{Sec-FinalComments}, where we also point out
our final comments.

\section{Cosmological set up \label{Sec-SetUp}}

This section presents the two cosmological models that are constrained by
observational data in this paper. The first one is a phenomenological model
that we call Unified Model (UM). The second one is the fiducial $\Lambda$CDM
model, considered here for the sake of comparison.

\subsection{Unified model \label{Sec-UM}}

Our framework will be a flat universe filled with baryonic matter and a
unified component of dark matter and dark energy. The Hubble function for this
model is%

\begin{equation}
H=H_{0}\sqrt{\Omega_{\text{U}}\left(  z,\alpha,\beta\right)  +\Omega
_{b0}\left(  1+z\right)  ^{3}}, \label{Eq:ModeloUnificado}%
\end{equation}
where $\Omega_{\text{U}}\left(  z,\alpha,\beta\right)  $ is the density
parameter of the unified fluid,%
\[
\Omega_{\text{U}}\left(  z\right)  =\Omega_{\text{U}0}\exp\left\{  \int
_{0}^{z}3\dfrac{\left[  1+w_{\text{U}}\left(  z^{\prime}\right)  \right]
}{1+z^{\prime}}dz^{\prime}\right\}  ,
\]
which is subjected to the constraint
\begin{equation}
\Omega_{\text{U}0}+\Omega_{b0}=1 \label{Eq:Vinculo1}%
\end{equation}
and depends on the redshift $z$\ and three free parameter $\Omega_{b0}$,
$\alpha$ and $\beta$ to be determined from adjustment to the available
observational data.

The parameter $w_{\text{U}}$ of the equation of state\ is a function of
$z$\ and describes the transition from the matter dominated dynamics to the
acceleration domination epoch. It is convenient to define \cite{Bruni2013A}%

\begin{equation}
w_{\text{U}}=\dfrac{1}{\pi}\arctan\left(  \alpha z-\beta\right)  -\dfrac{1}%
{2}. \label{Eq:weffMU}%
\end{equation}
The idea to propose a phenomenological parameterization which unifies the dark
components is not new. For instance, in the works \cite{Ribamar2007},
\cite{Ribamar2012} the authors use an expression exhibiting plots resembling
those built with Eq. (\ref{Eq:weffMU}); however, there is an important
conceptual difference between their reasoning and ours. Whereas in this paper
we adopt a dynamical approach, the authors of \cite{Ribamar2007},
\cite{Ribamar2012} use a kinematic one. The advantage of a
kinematic model in which one chooses to parameterize the deceleration
parameter $q$ in terms of the redshift $z$ -- as that of Ref.
\cite{Ribamar2012} -- is that very few assumptions on the nature of the dark
components are taken \textit{a priori}. On the other hand, dynamical models
parameterizing $w\left(  z\right)  $ are more physical in the sense that they
enable a meaningful perturbation theory (once they presuppose Einstein's
equation of gravity and standard cosmological assumptions).

Parameter $\alpha$\ gives the transition rate between the decelerated
expansion and the recent accelerated phase of the universe's evolution.
Parameter $\beta$\ provides the value for $w_{\text{U}}$\ today (null redshift).
Moreover,%

\begin{equation}
z_{\text{eq}}=\dfrac{\beta}{\alpha} \label{Eq:RedshiftTransicaoMU}%
\end{equation}
is the redshift corresponding to the equivalence between the dark energy and
the dark matter energy densities.\ This expression is obtained by taking
$w_{\text{U}}=-1/2$, the average of the values $w=0$ and $w=-1$.

\begin{figure*}[t]
\begin{center}
\subfloat[\hspace{-1.0cm} ]{ \label{Fig:compWeffa}
\includegraphics[width=0.5\textwidth]{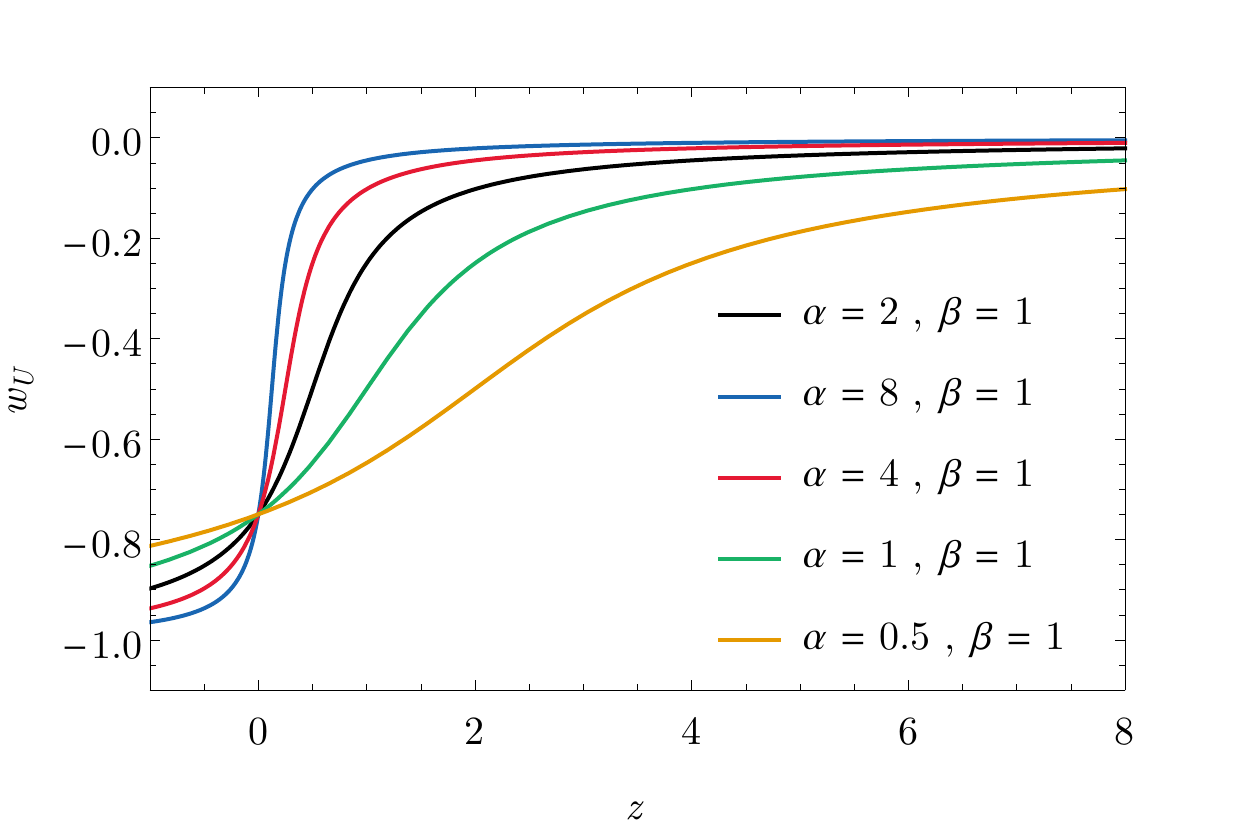}}
\subfloat[\hspace{-1.0cm} ]{ \label{Fig:compWeffb}
\includegraphics[width=0.5\textwidth]{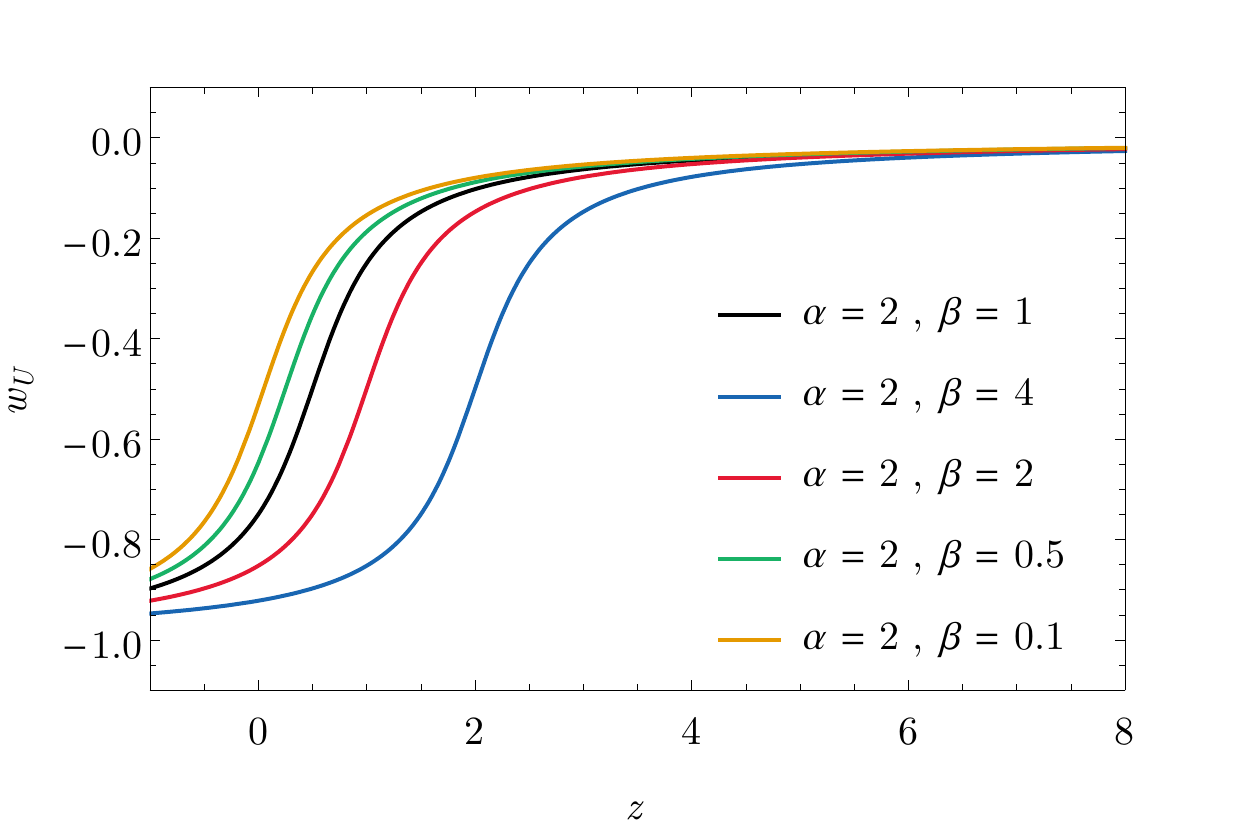}}
\end{center}
\caption{Curves of $w_{\text{\text{U}}}$ as a function of $z$\ for different
values of parameters $\alpha$ and $\beta$ --- see Eq. (\ref{Eq:weffMU}).}%
\label{Fig:compWeff}%
\end{figure*}

Fig. \ref{Fig:compWeffa} shows that the larger is $\alpha$ the greater is the
transition rate (if $\beta$ is kept constant). Fig. \ref{Fig:compWeffb}
illustrates the fact that the value of the redshift of equivalence grows with $\beta$
(for a given $\alpha$).

\subsection{$\Lambda$CDM}

We shall fit the concordance $\Lambda$CDM model to the observations using
the same data sets and techniques applied to our unified model for comparison.

The $\Lambda$CDM Hubble function for the flat universe is:%
\begin{equation}
H\left(  z\right)  =H_{0}\sqrt{(\Omega_{b0}+\Omega_{d0})\left(  1+z\right)
^{3}+\left(  1-\Omega_{b0}-\Omega_{d0}\right)  }, \label{Eq:ModeloLCDM}%
\end{equation}
where $\Omega_{b0}$\ is the density parameter for the baryonic matter and
$\Omega_{d0}$\ is the density parameter for the dark matter component. In the
$\Lambda$CDM cosmology, the constant $\Omega_{\Lambda}  = (1-\Omega_{b0}-\Omega_{d0})$\ is the density parameter of the dark energy, interpreted as a
cosmological constant. We define the effective equation of state parameter $w_{\text{dark}}$ for the $\Lambda $CDM model by the ratio of the pressure to the energy density of the dark components:
\begin{equation}
w_{\text{dark}}=-\frac{\Omega_{\Lambda}}{\Omega_{d0}\left(  1+z \right)  ^{3}+\Omega_{\Lambda}} \, .%
\end{equation}
Analogously to what we have done for the unified model, the equivalence redshift $z_{\text{eq}}$ is the solution to $w_{\text{dark}}(z_{\text{eq}})=-1/2$. Then:
\begin{equation}
z_{\text{eq}}=\left(  \dfrac{1-\Omega_{b0}-\Omega_{d0}}{\Omega_{d0}}\right)
^{1/3}-1 . \label{Eq:RedshiftTransicaoLC}%
\end{equation}

The above formulas will be useful in Sec. \ref{Sec-Data} when we obtain
parameters $\Omega_{b0}$\ and $\Omega_{d0}$\ using the observational data.


\section{Cosmological data sets, analysis and results \label{Sec-Data}}

The free parameters in Eqs. (\ref{Eq:ModeloUnificado}) and
(\ref{Eq:ModeloLCDM}) will be estimated using four different data sets:
Primordial Nucleosynthesis, Supernovae of the type Ia, Gamma-Ray Bursts
and Baryon Acoustic Oscillations.

\subsection{Primordial Nucleosynthesis data \label{Sec-PN}}

According to the Big Bang model, the nuclei of the light elements --- hydrogen
(H), deuterium (D), $^{3}$He, $^{4}$He e $^{7}$Li --- were created in the
first minutes of the Universe during a phase known as the primordial
nucleosynthesis \cite{Kirkman2003}. The abundances of these light elements
depend on the present-day value of the baryon density parameter $\Omega_{b0} $
and on the Hubble constant $H_{0}$ \cite{Pettini2012}. In fact, it is possible
to obtain $\Omega_{b0}h^{2}$ through a precise measurement of the primordial
abundance ratio for any two light nuclei species.

Among those nuclei formed during the primordial nucleosynthesis, the simplest
to be measure is the deuterium to hydrogen abundance ratio $\left(
D/H\right)  $ \cite{Omeara2006}. Ref. \cite{Adams1976} suggests to determine
this ratio using information from a special type of high-redshift quasar
(QSO), more specifically through damped Lyman alpha systems (DLA) spectra
\cite{Kirkman2003,Tytler1996,Burles1998,Omeara2001}%
.\footnote{We emphasize that only the QSO with the characteristics discussed
in \cite{Kirkman2003,Omeara2006,Pettini2008} can be used to
determine the abundance ratio $\left(  D/H\right)  $.} The deuterium to
hydrogen abundance ratio was given as $(D/H)=\left(  2.535\pm0.05\right) \times 10^{-5} $ by
Ref. \cite{Pettini2012} This result follows from the DLA QSO SDSS J1419+0829
spectrum. The above value for $(D/H)$\ leads to%
\begin{equation}
\Omega_{b0}h^{2}=0.0223\pm0.0009~. \label{Eq:Barion_h^2}%
\end{equation}

Refs. \cite{Riess2011} and \cite{Freedman2012} discuss measurements of the
Hubble constant $H_{0}$ with a negligible dependence on the cosmological
model. These two sources enable one to obtain the normalized Hubble constant
$h$ in Eq. (\ref{Eq:Barion_h^2}) as:%
\begin{equation}
\left\{
\begin{array}
[c]{l}%
h_{\text{R}}=0.738\pm0.028\\
h_{\text{F}}=0.743\pm0.015(\text{sta})\pm0.021\left(  \text{sys}\right)
\end{array}
\right.
\begin{array}
[c]{l}%
\text{(Riess);}\\
\text{(Freedman).}%
\end{array}
\label{Eq:CteHubble}%
\end{equation}
On the order hand, the $H_{0}$ measured for Planck satellite \cite{Planck2013}
indicates $h=0.673\pm0.012$. So, there is a noticeable $2.5\sigma$\ discrepancy
between the values for\ $h$\ given by Riess and Freedman\ and the one measured
by Planck satellite. In this work, we shall adopt a conservative stance and use
$h=0.74$. Nevertheless we use $\sigma_{h}=0.07$ as uncertainty in order to
accommodate Planck's value with a confidence interval of $1\sigma$. Using the
data in (\ref{Eq:CteHubble}) and Eq. (\ref{Eq:Barion_h^2}), one can estimate
the baryon density parameter as $\Omega_{b0}^{\,\text{PN}}=\left(  0.0407\pm0.0079\right)  $.

\subsection{SNIa data\label{Sec-SNIa}}

The supernovae are super-massive star explosions with intense luminosity. Among
them, type Ia supernovae (SNIa) are the most important for cosmology since
they can be taken as standard candles due to their characteristic luminosity curves.

In order to estimate the cosmological parameters of the unified model, we will
employ the 580 SNIa compilation available in Ref. \cite{Union21} by the
Supernova Cosmology Project (SCP).\footnote{Union2.1 data set, including the
580 supernovae, is available at the electronic address
\url{http://supernova.lbl.gov/Union}.} Union2.1 data set presents the
redshift $z$\ of each supernova and the related distance modulus $\mu$
accompanied by its uncertainty $\sigma_{\mu}$.

The distance modulus $\mu$ is a logarithmic function of the normalized luminosity distance
$d_{h}$:
\begin{equation}
\mu (  z;\vec{\theta},\mathcal{M} ) = 5\log d_{\text{h}} (
z;\vec{\theta} )  +\mathcal{M\,}, \label{Eq.Modulo}
\end{equation}
with
\begin{equation}
d_{\text{h}}\equiv\dfrac{H_{0}}{c}d_{\text{L}}=\left(  1+z\right) H_{0} \int_{0}^{z}\dfrac{dz^{\prime}}{H (  z^{\prime};\vec{\theta}\,)  } \, . \label{Eq:dh}
\end{equation}
$\mathcal{M}$ is a constant depending on the Hubble
constant $H_{0}$, the speed of light $c$ and the absolute magnitude of the
standard supernova in the regarded band \cite{Rosenfeld2006}. $\vec{\theta}$
is the vector of parameters for the particular cosmological model under
consideration. We shall not discuss the quantities encapsulated in
$\mathcal{M}$ since they are not of our concern here; in fact, $\mathcal{M}%
$\ is marginalized in the statistical treatment of the data. In fact, we define
\begin{equation}
\chi_{\text{S+N}}^{2} (  \vec{\theta}\, )  =\chi_{\text{SN,}m}%
^{2} (  \vec{\theta} \, )  +\chi_{\text{PN}}^{2}~ \label{Chi2_SN_PN}%
\end{equation}
where 
\begin{equation}
\chi_{\text{PN}}^{2}=\dfrac{\left(  \Omega_{b0}-\Omega_{b0}^{\,\text{PN}%
}\right)  ^{2}}{\left(  \sigma_{\Omega_{b0}^{\,\text{PN}}}\right)  ^{2}}~,
\label{Eq:DensBarions}%
\end{equation}
and the function $\chi_{\text{SN,}m}^{2}$\ comes from the $\chi_{\text{SN}%
}^{2}$ of Union2.1 supernovae data,%
\begin{equation}
\chi_{\text{SN}}^{2} (  \vec{\theta},\mathcal{M} )  =\sum_{i=1}%
^{580}\dfrac{\left[  \mu_{i}-5\log d_{h} (  z_{i};\vec{\theta} \,)
-\mathcal{M}\right]  ^{2}}{\sigma_{\mu_i}^{2}}, \label{eq:NS_Chi2}%
\end{equation}
after analytic marginalization of the parameter $\mathcal{M}$
\cite{Goliath2001}.

\begin{figure}

\includegraphics[width=0.5\textwidth]{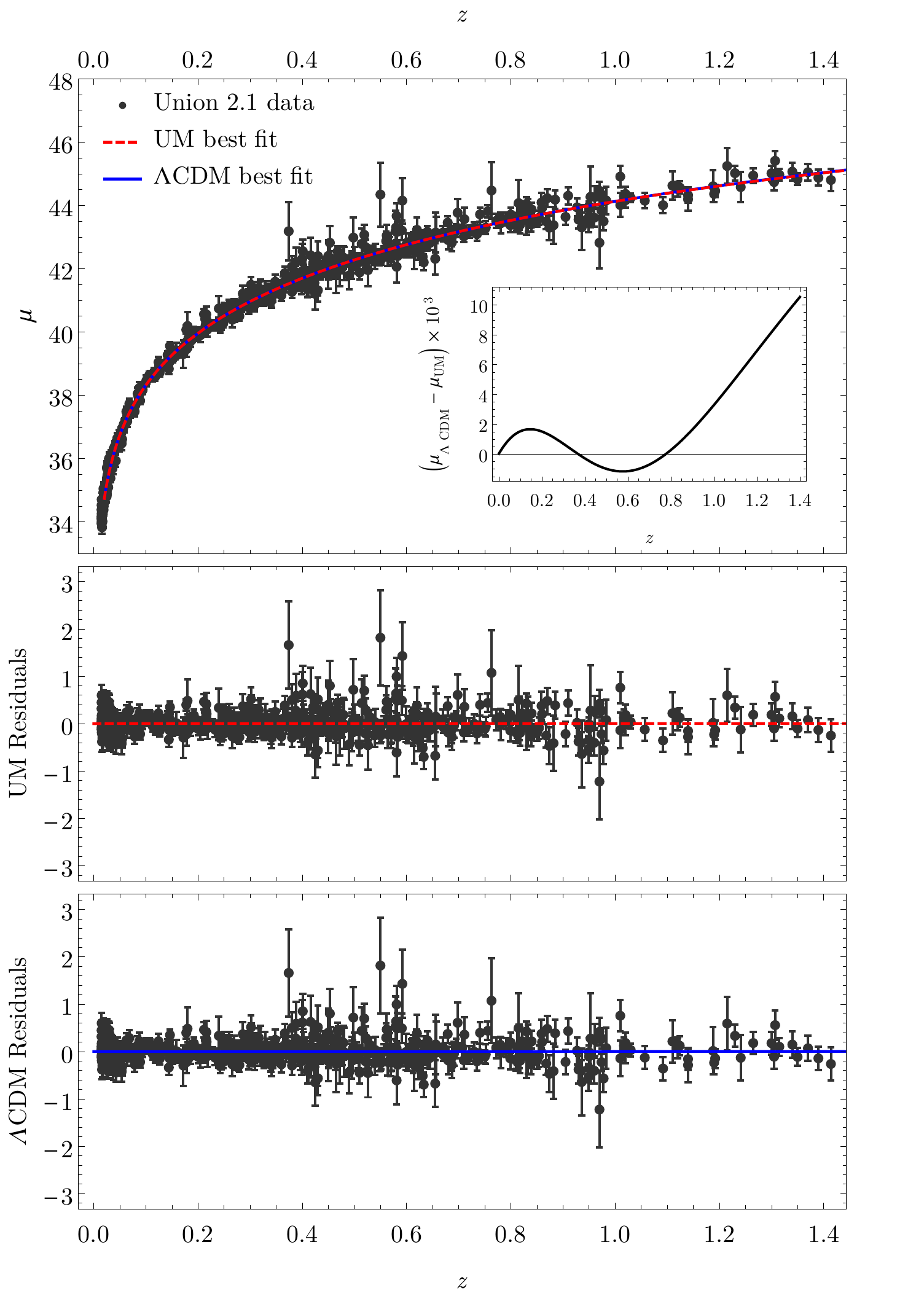}

\caption{SNIa data and the distance modulus curves built with the best-fit values of UM and $\Lambda$CDM (top). The curve $\mu(z)$ for UM almost coincides with the one for $\Lambda$CDM; the small differences between $\mu_{\text{UM}}$ and $\mu_{\Lambda \text{CDM}}$ are emphasized in the plot of $(\mu_{\Lambda \text{CDM}} - \mu_{\text{UM}})\times 10^{3}$ as a function of $z$. The residual plots for UM (center) and $\Lambda$CDM (bottom) are also show.}
\label{Fig:Residuals}
\end{figure}

\begin{figure*}[t]
\subfloat[Unified Model]{ \label{Fig:GrafManchaUNa}
\includegraphics[width=0.5\textwidth]{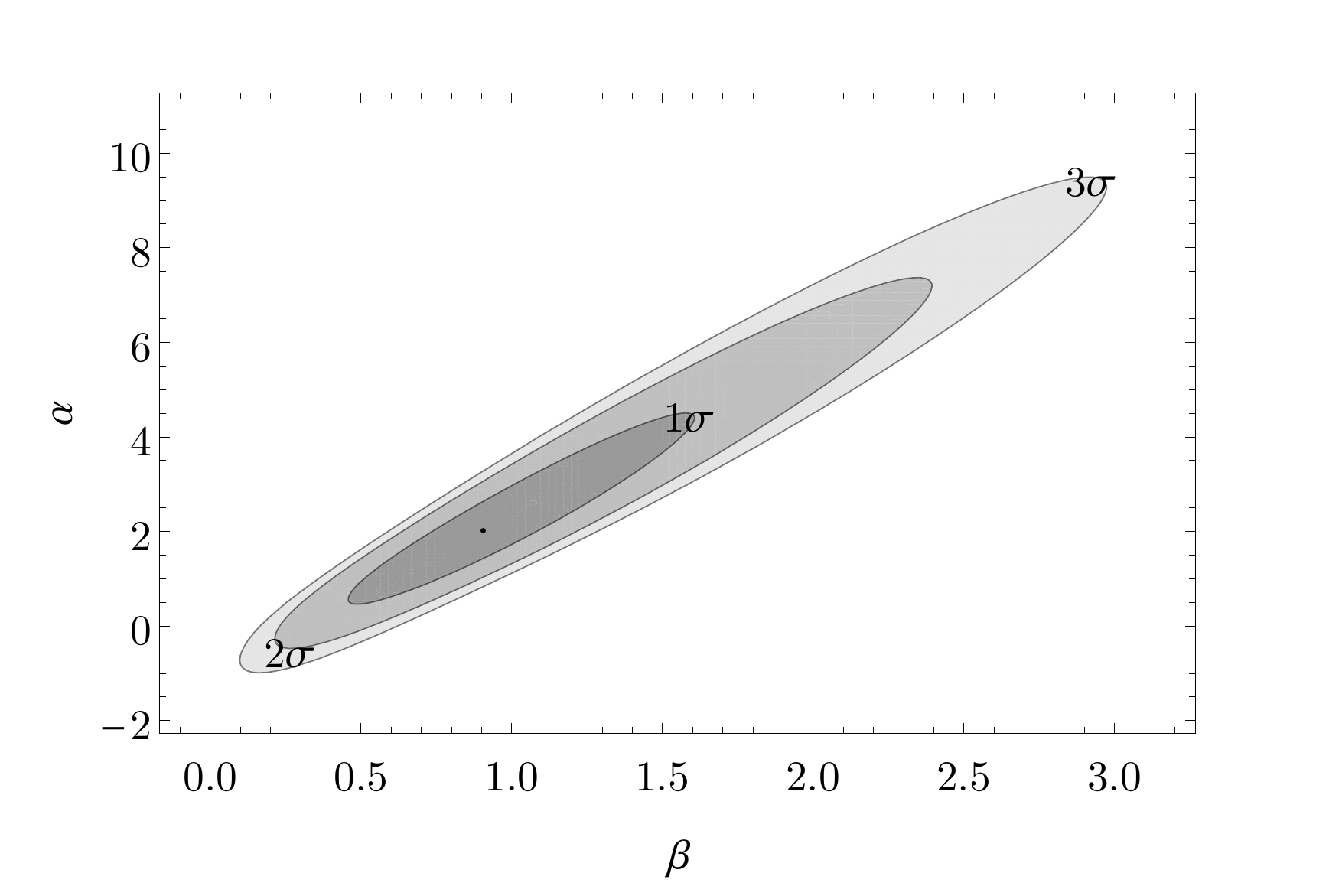}}
\subfloat[ $\Lambda$CDM ]{ \label{Fig:GrafManchaUNb}
\includegraphics[width=0.5\textwidth]{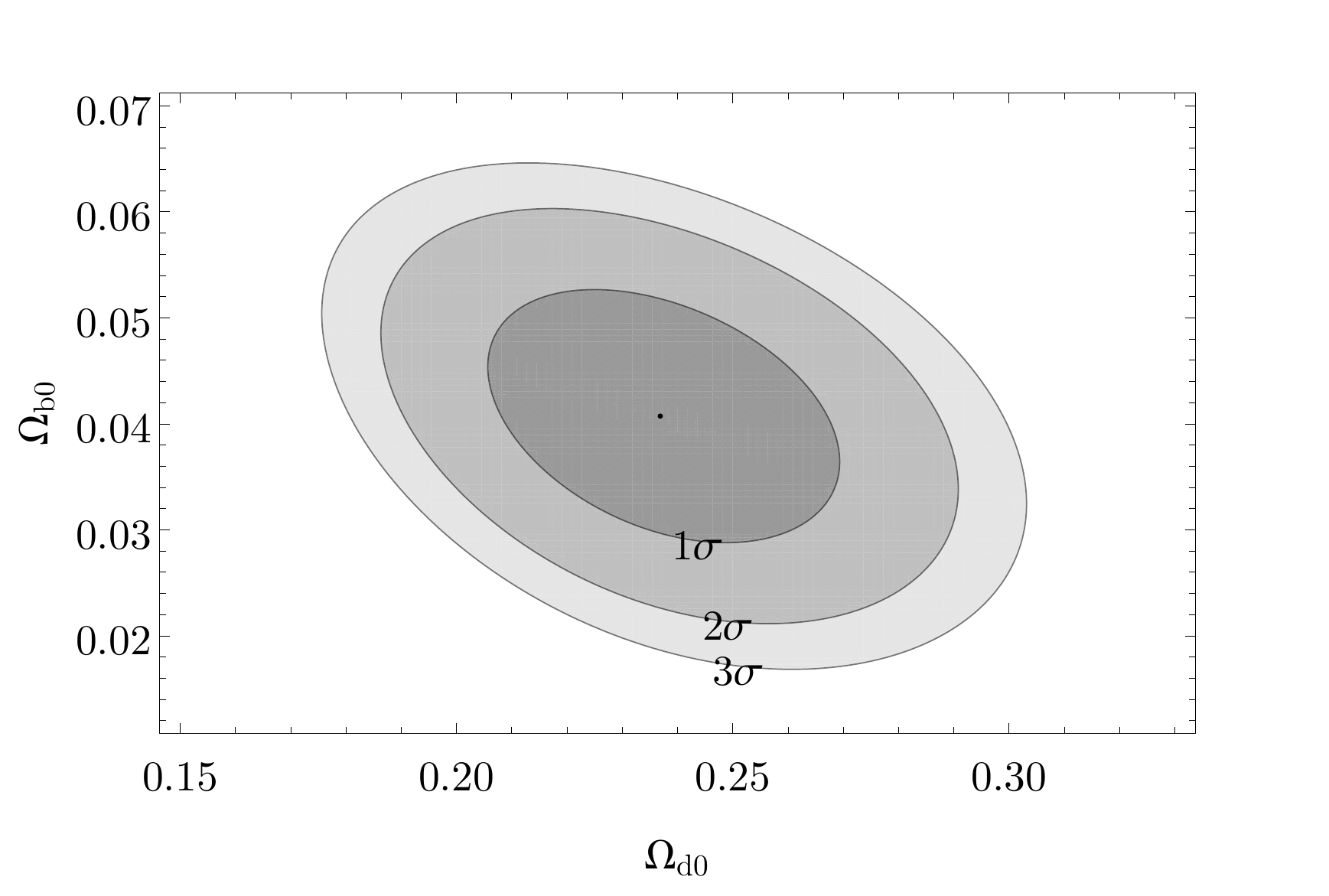}}

\caption{(a) UM: 68.3\%, 95.4\% and 99.7\% confidence regions of the $(\beta , \alpha)$ plane from SNIa data combined with the constraint from PN. (b) $\Lambda$CDM model: 68.3\%, 95.4\% and 99.7\% confidence regions of the $(\Omega_{d0} , \Omega_{b0})$ plane from SNIa data combined with the constraint from PN. Confidence contours include statistical and systematic errors in both cases.}%
\label{Fig:NS_GrafMancha}%
\end{figure*}

We estimate the vector of parameters $\vec{\theta}_{\text{UM}}=\left(  \Omega_{b0},\alpha,\beta\right)  $ and $\vec{\theta}_{\Lambda\text{CDM}}$\ $=\left(  \Omega_{b0},\Omega_{d0}\right) $ by minimizing $\chi_{\text{S+N}}^{2} ( \vec{\theta} ) $ for the UM and $\Lambda$CDM model. The values
of the parameters for both models are found in Table\ \ref{Tab:Parametros}. The best-fit parameters are used to build distance moduli curves for both models. The upper part of Fig. {\ref{Fig:Residuals}} show how well UM vs. $\Lambda$CDM fit the data. The residual plots are at the lower part of Fig. {\ref{Fig:Residuals}}. Confidence region graphs (with $1\sigma$, $2\sigma$\ and $3\sigma$) are displayed in Fig.
\ref{Fig:NS_GrafMancha}.

\subsection{GRB data\label{subsec:CalibrarGRB}}

The SNIa data provide us with reliable cosmological information till redshifts
of the order of $1.7$ (cf. \cite{z1.7a}). On the other hand, cosmic microwave
background anisotropy measurements permit us to access information about the
large scale universe at $z\sim1000$ \cite{Huz1000}. In between, there is a
large redshift interval observationally inaccessible; scientific community is
making great effort to collect astronomical data to fill in this gap. Perhaps the
most promising candidates for this scope are the Gamma-Ray Bursts. It is expected that a
fraction $\gtrsim50\%$\ of the observed GRB have $z>5$\ and the redshift values
of these objects may be as large as $10$ or even greater \cite{Bromn}.

Even though we do not fully understand GRB emission mechanism, they are
considered excellent candidates to standard candles because of their intense
brightness \cite{Wei2}, \cite{Li1a}. That is the reason why many authors have
been proposing empirical luminosity correlation functions that standardize GRB
as distance indicators \cite{Amati1,Guirlanda1,Liang1,Firmani1}.

An additional problem to the use of GRB is the so called circularity problem.
Unlike what happens in the supernovae case, there is no data set that is
completely model independent and which could be used to calibrate GRB
distance curves \cite{Wei2}, \cite{Schaefer}. A number of different
statistical methods were suggested to overcome this model dependence; e.g. see
Refs.  \cite{Wei2, Wei3, Graziani,Wang1,Guirlanda3,Ghirlanda2,Li2,Liang2,Liang3}.

This work make use of the 138 Gamma-Ray Bursts compiled in Ref. \cite{Liu2014b}. They were calibrated according to the method described in Ref. \cite{Wei3},
which tries to eliminate model dependence. Two different groups of GRB were
considered: the low-redshift set has $z<1.4$; the high-redshift one presents
events with $z>1.4$.

\begin{figure*}[t]
\subfloat[\hspace{-1.0cm} ]{ \label{Fig:CalibrGRBa}
\includegraphics[width=0.47\textwidth]{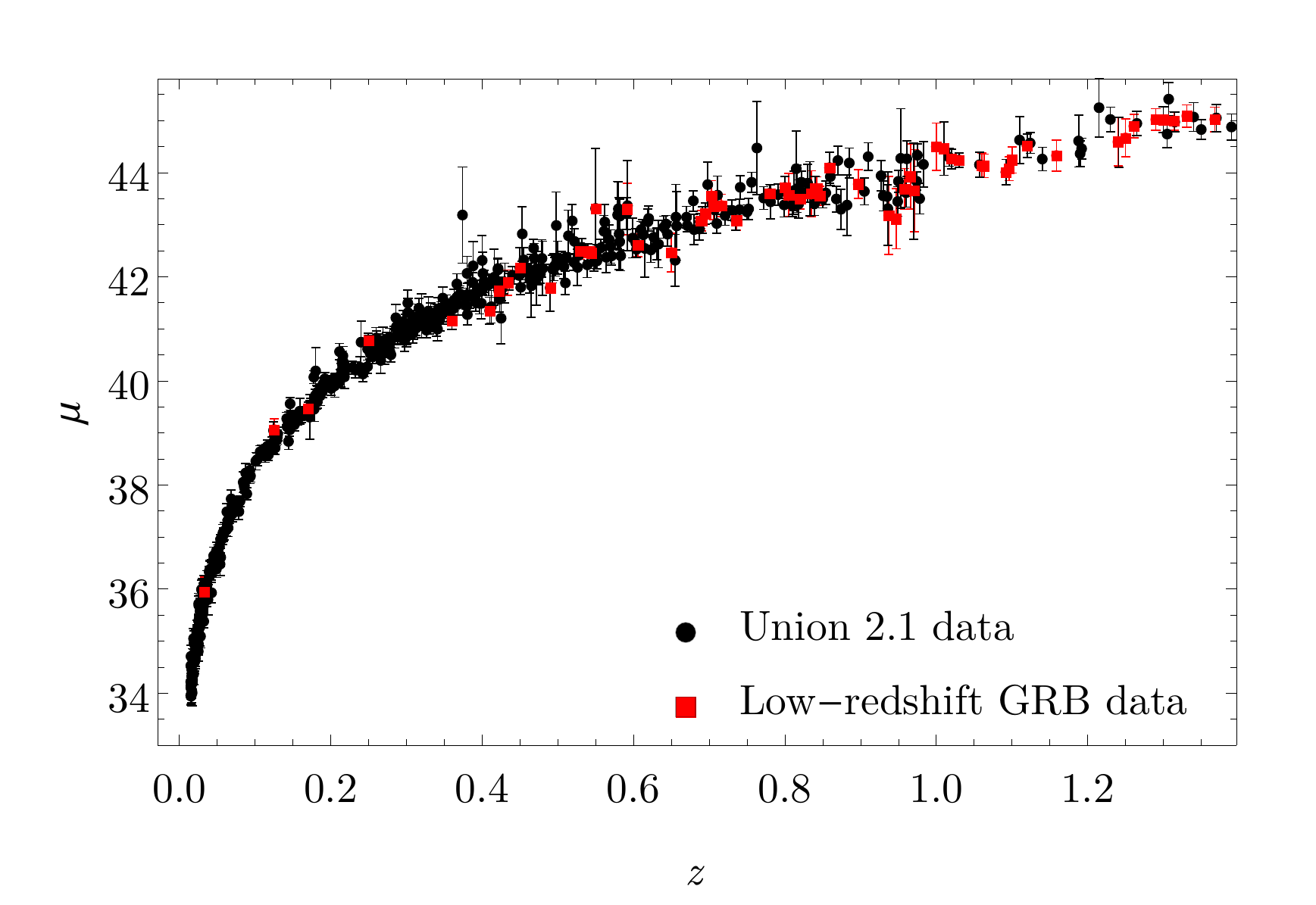}} \hspace{5mm}
\subfloat[\hspace{-1.0cm} ]{ \label{Fig:CalibrGRBb}
\includegraphics[width=0.47\textwidth]{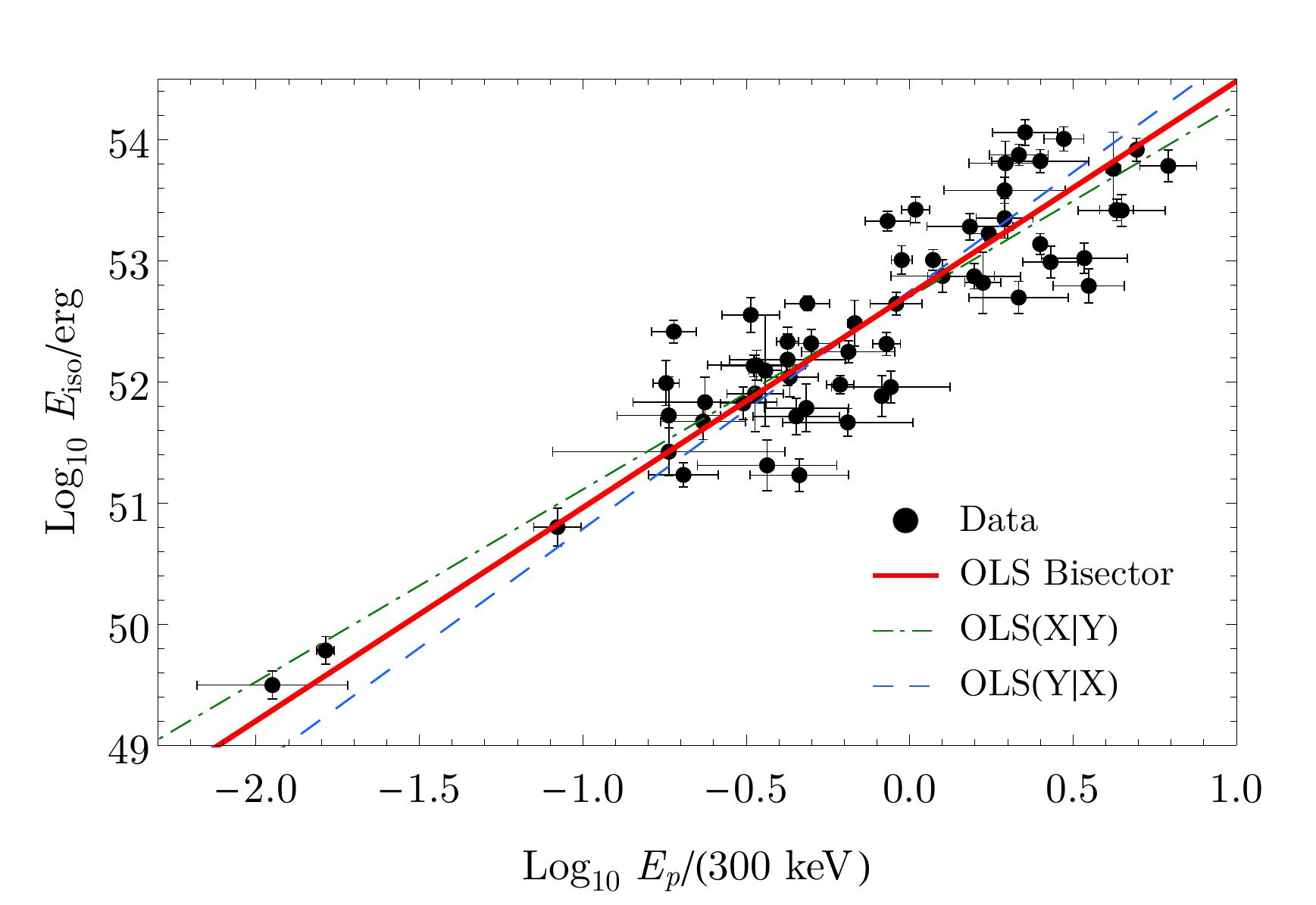}}

\caption{(a) $\mu\left(  z\right)  $ points for the Union2.1 SNIa data set (in
black) and low-redshift GRB data (red). (b) Linear regression procedures for
determining parameters $b$ and $\lambda$ in the relation by Amati.}%
\label{Fig:CalibrGRB}%
\end{figure*}

The distance modulus of the low-redshift GRB were determined using the SNIa
data in the following way. We built the plot of the distance modulus versus
the redshift for the 580 supernovae of the Union2.1 data set. The supernovae
with the same redshift had their distance modulus values averaged. The points
in the plot $z\times\mu$ were interpolated to provide a function $\mu
=\mu\left(  z\right)  $\ with domain $0\leq z\leq1.4$. There are many
interpolation techniques such as linear, cubic and Akima's interpolations
\cite{Akima}. We used these three methods and chose the last one for building
the function $\mu\left(  z\right)  $\ because Akima's technique is the one
giving a curve that intercepts the points in a more smooth and natural way --
see Appendix \ref{Akima}. With these SNIa low-redshift $\mu\left(  z\right)
$\ it is possible to estimate the distance modulus of each one of the 59
low-redshift GRB. These values of $\mu$\ are then substituted in%
\begin{equation}
\mu=5\log_{10}\dfrac{d_{\text{L}}}{\text{Mpc}}+25 \label{Eq:ModGRB}%
\end{equation}
to give the associated luminosity distances $d_{L}$. They, in turn, appear in
the expression for the isotropically radiated equivalent energy:%
\begin{equation}
E_{\text{iso}}=\dfrac{4\pi S_{\text{bolo}}d_{L}^{2}}{1+z}, \label{Eq:Eiso}%
\end{equation}
where $S_{\text{bolo}}$\ is the GRB observed bolometric fluency. In the work
\cite{Amati1}, Amati noticed the correlation between the energy peak of the
GRB spectrum ($E_{\text{p}}$) and the isotropically radiated energy
($E_{\text{iso}}$), formulating the equation:
\begin{equation}
\log_{10}\dfrac{E_{\text{iso}}}{\text{erg}}=\lambda+b\log_{10}\dfrac
{E_{\text{p}}}{300~\text{keV}}, \label{Eq:logEiso}%
\end{equation}
which is known as the Amati's relation. We determine parameters $\lambda$ and
$b$ by using the low-redshift GRB data set, with the $E_{\text{p}}$ data
available in Ref. \cite{Liu2014b} and the $E_{\text{iso}}$\ obtained from the
SNIa calibration curve. Parameters $\lambda$ and $b$\ are obtained from a
linear fit to the Amati's relation. The usual linear fit procedures in
astronomy are the ordinary least-squares regression of the dependent variable
$Y$ against the independent variable $X$\ -- OLS(Y$\vert$X) -- and the ordinary least-squares of $X$ on $Y$\ -- OLS(X$\vert$Y). However, if there is a domain within which occurs an intrinsic scattering
of the data with respect to the individual uncertainties, it is preferable to
use the OLS bisector method, as described in Ref. \cite{LinearRegr}. Following
the procedure in this reference, we performed linear regressions using the
three methods above; the values obtained for the parameters $b$ and $\lambda$
of the Amati's relation are displayed in Table \ref{Tab:ParametrosAjuste}; the
straight lines built from those parameters are shown in Fig.\ref{Fig:CalibrGRB}.

\begin{table}[tbh]
\caption{Parameter $b$ and $\lambda$ of the Amati's relation.}%
\label{Tab:ParametrosAjuste}%
\centering$%
\begin{tabular}
[c]{ccccc}\hline
Method & $b$ & $\sigma_{b}$ & $\lambda$ & $\sigma_{\lambda}$\\\hline
\multicolumn{1}{l}{OLS(X$|$Y)} & $1.564$ & $0.084$ & $52.74$ & $0.06$\\
\multicolumn{1}{l}{OLS(Y$|$X)} & $1.861$ & $0.099$ & $52.79$ & $0.06$\\
\multicolumn{1}{l}{OLS bisector} & $1.703$ & $0.053$ & $52.77$ &
$0.06$\\\hline
\end{tabular}
\ $\end{table}

We decided to adopt the values of $\lambda$ and $b$ given by the OLS
bisector\ method once the intrinsic dispersion of the data is dominant over
the observational errors. Then, we calculated the quantity $\log_{10}E_{\text{iso}}
$\ for the high-redshifts GRB and their distance modulus%
\begin{equation}
\mu=\dfrac{5}{2}\log_{10}\dfrac{E_{\text{iso}}}{\operatorname{erg}
}+\dfrac{5}{2}\log_{10}\dfrac{\left(  1+z\right)  }{4\pi S_{\text{bolo}}}+25
\label{Eq:ModEiso}
\end{equation}
with an associated uncertainty
\begin{equation}
\sigma_{\mu}=\sqrt{\left(  \dfrac{5}{2}\sigma_{\log_{10}E_{\text{iso}}%
}\right)  ^{2}+\left(  \dfrac{5}{2\ln10}\dfrac{\sigma_{S_{\text{bolo}}}%
}{S_{\text{bolo}}}\right)  ^{2}}. \label{Eq:IncModEiso}%
\end{equation}
The uncertainty related to $\log_{10}E_{\text{iso}}$\ is given by:%

\begin{eqnarray}
\sigma_{\log_{10}E_{\text{iso}}}^{2}&=&\sigma_{\lambda}^{2}+ \left(
\dfrac{b}{\ln10}\dfrac{\sigma_{E_{\text{p}}}}{E_{\text{p}}}\right)
^{2}+\nonumber \\
 & &+\left(  \sigma
_{b}\log_{10}\dfrac{E_{\text{p}}}{300~\text{keV}}\right)  ^{2}+\sigma_{E_{\text{sys}}}^{2}. \label{Eq:log10Eiso}
\end{eqnarray}
This equation is obtained from Amati's relation through error propagation. We
also added the contribution of the systematic error $\sigma_{E_{\text{sys}}}%
$\ coming from extra dispersion in the luminosity relations. This systematic
error is a free parameter and can be estimated by imposing $\chi_{\text{red}%
}^{2}=1$ on the curve fitting to the luminosity plots. This was done in Ref.
\cite{Schaefer}, and the value obtained is: $\sigma_{E_{\text{sys}}}^{2}=0.39$.

After performing the GRB calibration using Union2.1 SNIa data, one obtains a set
of values for the distance modulus $\mu\left(  z\right)  $ (and its
uncertainty $\sigma_{\mu}$) for 79 high-redshift GRB. This set of values for
$\mu\left(  z\right)  \pm\sigma_{\mu}$ is shown in Appendix
\ref{App:TabelaHighGRB}. It is used to build the function $\chi_{\text{GRB}}^{2}$:
\begin{equation}
\chi_{\text{GRB}}^{2} (  \vec{\theta},\mathcal{M} )  =\sum_{i=1}%
^{79}\dfrac{\left[  \mu_{i}-5\log \frac{d_{L} (  z_{i};\vec{\theta} \, )}{\text{Mpc}}
- 25 \right]  ^{2}}{\sigma_{\mu_i}^{2}}.
\label{eq:NSG_Chi2}
\end{equation}

Now we use as input to our statistical treatment the three sets
of data discussed so far (Union2.1 data; the value of $\Omega_{b0}^{\,\text{PN}}%
$\ coming from PN; and, the  79 high-redshift GRB data duly calibrated) to estimate
the cosmological parameters. 

\begin{figure*}[t]

\subfloat[Unified Model]{ \label{Fig:GrafManchaUNBa}
\includegraphics[width=0.5\textwidth]{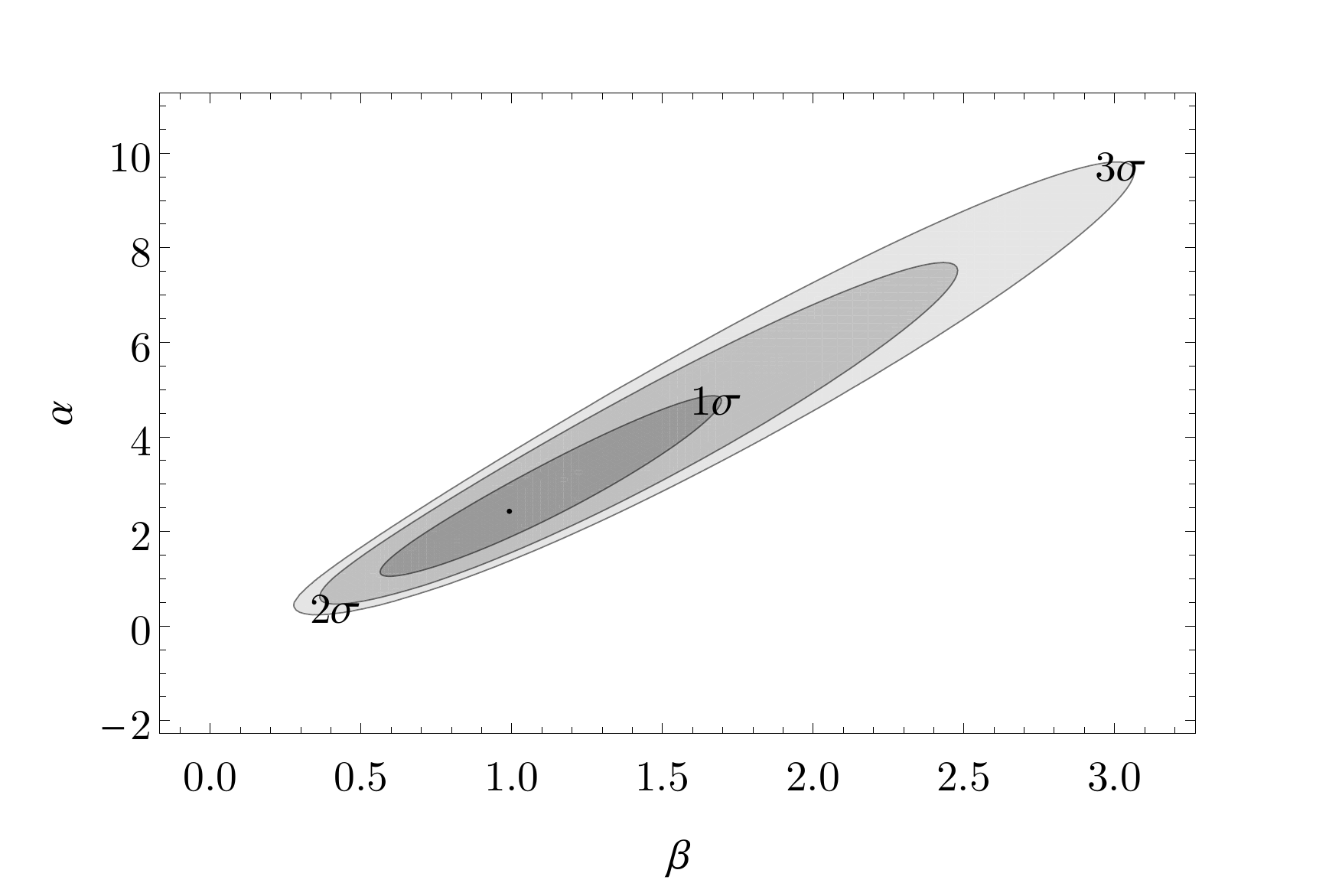}}
\subfloat[ $\Lambda$CDM ]{ \label{Fig:GrafManchaUNBb}
\includegraphics[width=0.5\textwidth]{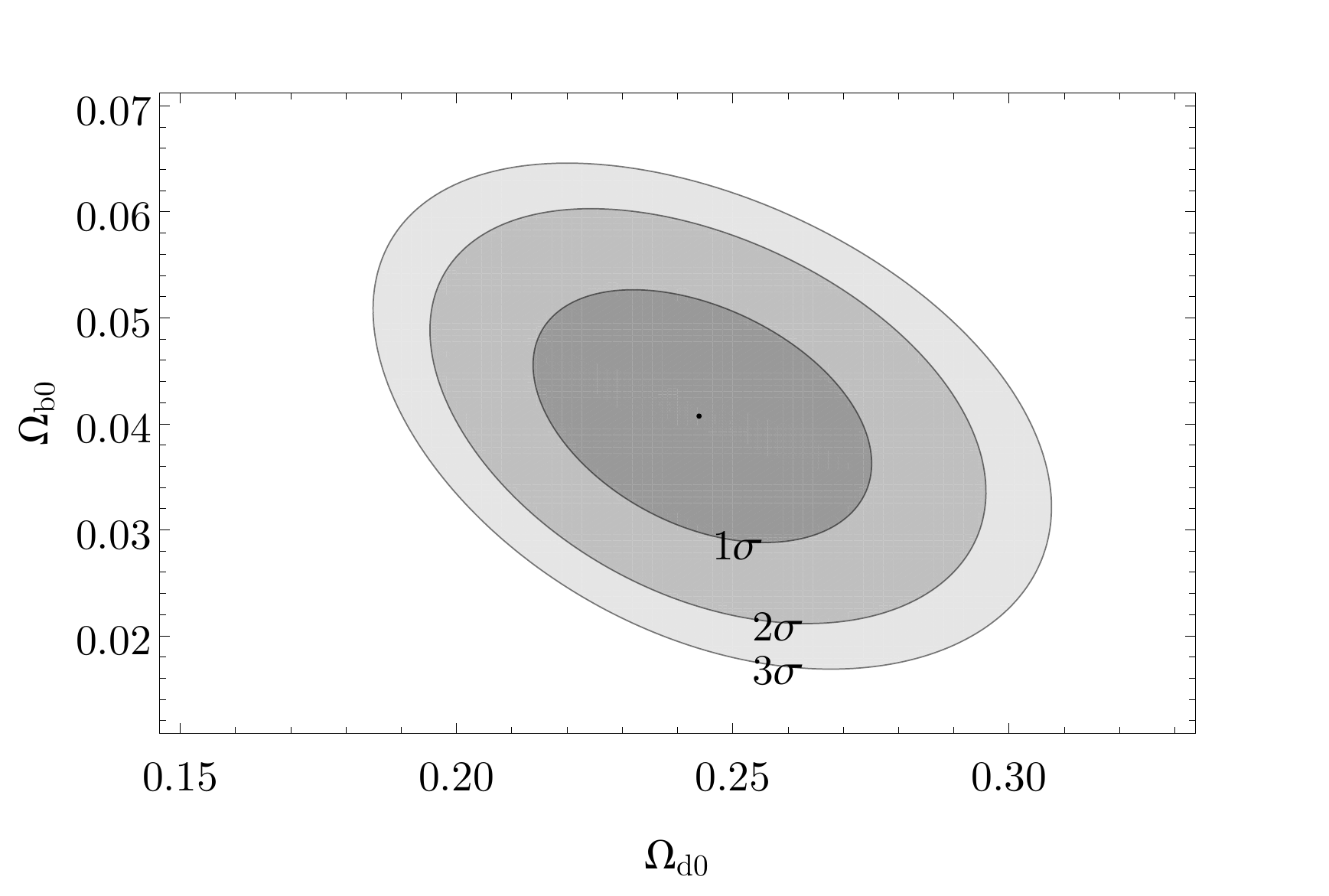}}

\caption{(a) UM: 68.3\%, 95.4\% and 99.7\% confidence regions of the $(\beta , \alpha)$ plane from SNIa and PN plus high-redshift GRB data. (b) $\Lambda$CDM model: 68.3\%, 95.4\% and 99.7\% confidence regions of the $(\Omega_{d0} , \Omega_{b0})$ plane from SNIa data combined with PN and GRB data. Confidence contours include statistical and systematic errors in both cases.}
\label{Fig:NSG_GrafMancha}%
\end{figure*}

The best-fit values and single-parameter estimates are displayed in Table \ref{Tab:Parametros}. Fig. \ref{Fig:NSG_GrafMancha} exhibits the
double-parameter estimates with $1\sigma$, $2\sigma$\ e $3\sigma$ confidence
regions. 

\subsection{BAO data\label{Sec-BAO}}

Before the last scattering, the baryon-photon plasma weakly coupled oscillated
due to a competition between the gravitational collapse and the radiation
pressure \cite{Ellis2012}. According to \cite{Hu1996}, the velocity of the
resulting sound waves in the plasma is $c_{s}=1/\sqrt{3(1+3\rho_{b}%
/4\rho_{\gamma})}$. The stagnation of these waves after the decoupling lead to
an increase of the baryon density at the scales corresponding to the distance
covered by the acoustic wave until the decoupling time. This effect produces a
peak of baryon acoustic oscillation (BAO) in the galaxy correlation function.
BAO peaks data present very small systematic uncertainties when compared to
the other cosmological data sets \cite{Percival2009,Albrecht2006b}. This is clearly an
advantage to be used.

The baryon release marks the end of the Compton drag epoch and occurs at the
redshift $z_{\text{drag}}\simeq1059$ \cite{Planck2013}. The sound horizon $r_{s}$
determines the location of the length scale of the BAO peak. It is given by:%

\begin{equation}
r_{s}\left(  z_{\text{drag}}\right)  =\dfrac{1}{\sqrt{3}}\int_{z_{\text{drag}%
}}^{\infty}\dfrac{c~dz}{\bar{H}\left(  z\right)  \sqrt{1+\tfrac{3}{4}\tfrac
{1}{\left(  1+z\right)  }\tfrac{\Omega_{b0} }{\Omega_{\gamma0}}}}.
\label{Eq:rs}%
\end{equation}
The original Hubble function of the unified model,
Eq.(\ref{Eq:ModeloUnificado}), must be modified to%
\begin{eqnarray}
\bar{H}\left(  z\right) &=& H_{0} \Big[\Omega_{\text{U}}\left(  z,\alpha
,\beta\right) + \nonumber  \\
 & &+\Omega_{b0}\left(  1+z\right)  ^{3}+\Omega_{r0}\left(
1+z\right)^{4} \Big]^{1/2}, \label{Eq:ModRad}%
\end{eqnarray}
in order to include the radiation-like term $\Omega_{r0}\left(  1+z\right)
^{4}$. This is necessary here because we are dealing with the $z>1000$,
corresponding to the baryon-photon decoupling epoch, when the radiation was by
no means negligible. The fact that $\lim_{z\rightarrow0}\bar{H}\left(
z\right)  =H\left(  z\right)  $\ guarantees that $\bar{H}\left(  z\right)
$\ describes the same unified model we have been discussing from the beginning
of the paper.

For the sake of comparison, we shall study the sound horizon $r_{s}$ for the
$\Lambda$CDM model. The Hubble function for this case is:%
\begin{eqnarray}
\bar{H}\left(  z\right) &=& H_{0} \Big[\left(  1-\Omega_{b0}-\Omega_{d0}-\Omega_{r0}\right)  + \nonumber  \\
 & &+\left(  \Omega_{b0}+\Omega_{d0}\right) \left(
1+z\right)  ^{3}+\Omega_{r0}\left(
1+z\right)^{4} \Big]^{1/2}.\nonumber \\ \label{H bar}%
\end{eqnarray}

The density parameter $\Omega_{r0}$ describes the contributions from the
photons as well as that from the ultra-relativistic neutrinos. In accordance
with \cite{Komatsu2009,Ichikawa1},%
\begin{equation}
\Omega_{r0}=\Omega_{\gamma0}\left(  1+0.2271N_{\text{eff}}\right)  ,
\label{Omega r 0}%
\end{equation}
where $N_{\text{eff}}=3.046$ is the effective number of neutrinos.\ The
present-day value of the photon density parameter is $\Omega_{\gamma
0}=5.46\times10^{-6}$, cf. Ref. \cite{PDG}.

When we substitute (\ref{Eq:ModRad}) into (\ref{Eq:rs}), the sound horizon
turns out to be a function of the free parameters $\alpha$\ and $\beta
$\ present in our unified model: $r_{s}=r_{s}\left(  \alpha,\beta\right)  $.
The sound horizon is then used to constraint $\alpha$\ and $\beta$. This is
done in the following way. BAO data allow us to obtain the angular diameter
distance $D_{A}\left(  z\right)  $, achieved from the observation of the
clustering perpendicular to the line of sight, and the Hubble function
$H\left(  z\right)  $, measured through the clustering along the line of
sight. However, $D_{A}\left(  z\right)  $\ and $H\left(  z\right)  $\ are not
obtained independently, but through the distance scale ratio \cite{Farook}%
\begin{equation}
d_{z}=\dfrac{r_{s} (  z_{\text{drag}},\vec{\theta} \,)  }{D_{v} (
z,\vec{\theta}\,)  }, \label{Eq:dz}%
\end{equation}
where%
\begin{equation}
D_{v} (  z,\vec{\theta} \,)  =\left[  \left(  1+z\right)  ^{2}D_{A}^{2} (  z,\vec{\theta} \,)  \dfrac{cz}{H (  z^{\prime},\vec{\theta} \,)  }\right]  ^{1/3} \label{Eq:Dv}
\end{equation}
is the effective distance ratio, and%
\begin{equation}
D_{A} (  z,\vec{\theta}\,)  =\dfrac{1}{\left(  1+z\right)  }\int
\dfrac{cdz^{\prime}}{H (  z^{\prime},\vec{\theta}\,)  }~.
\label{Eq:Da}%
\end{equation}

We perform a data fit to the three values measured for the distance scale ratio
$d_{z}$ --- see Table \ref{Tab:DadosBAO}. These are non-correlated BAO peaks data.

\begin{table}[tbh]
\caption{BAO data.}%
\label{Tab:DadosBAO}%
\centering%
\begin{tabular}
[c]{cccc}\hline
Survey & $z$ & $d_z$ & Reference\\\hline
6dFGS & $0.106$ & $0.336\pm0.015$ & \cite{Beutler2011}\\
Boss & $0.32$ & $0.1181\pm0.0023$ & \cite{Anderson2014}\\
Boss & $0.57$ & $0.07261\pm0.00071$ & \cite{Anderson2014}\\\hline
\end{tabular}
\end{table}

Function $\chi_{\text{BAO}}^{2}$,%
\begin{equation}
\chi_{\text{BAO}}^{2} (  \vec{\theta}\,)  =\sum_{i=1}^{3}\dfrac
{1}{\sigma_{i}^{2}}\left[  d_{z,i}-\dfrac{r_{s} (  z_{\text{drag}},\vec{\theta} \,)}{D_{v} (  z_{i},\vec{\theta} \,)  }\right]^{2}~, \label{eq:NSB_Chi2}
\end{equation}
is calculated using the data in Table \ref{Tab:DadosBAO}. We add to this
function the expression for $\chi_{\text{S+N}}^{2}$, Eq. (\ref{Chi2_SN_PN}),
so that we take into account BAO data together with SNIa and PN\ data sets. By minimizing the complete $\chi^2$, one finds the best-fit values and the single-parameter estimates shown in Table \ref{Tab:Parametros}. Fig. \ref{Fig:NSB_GrafMancha} displays the confidence regions related to the two-parameter estimates.

According to the values shown in
Table \ref{Tab:Parametros}, the set including SNIa, PN and BAO is rather restrictive in comparison with
the results obtained with SNIa and PN only. By considering the
BAO peaks in the statistical treatment we reduced considerably the $1\sigma
$-confidence interval of the single-parameter estimates. 

\begin{widetext}

\begin{figure*}[t]

\subfloat[Unified Model]{ \label{Fig:GrafManchaUNBAOa}
\includegraphics[width=0.5\textwidth]{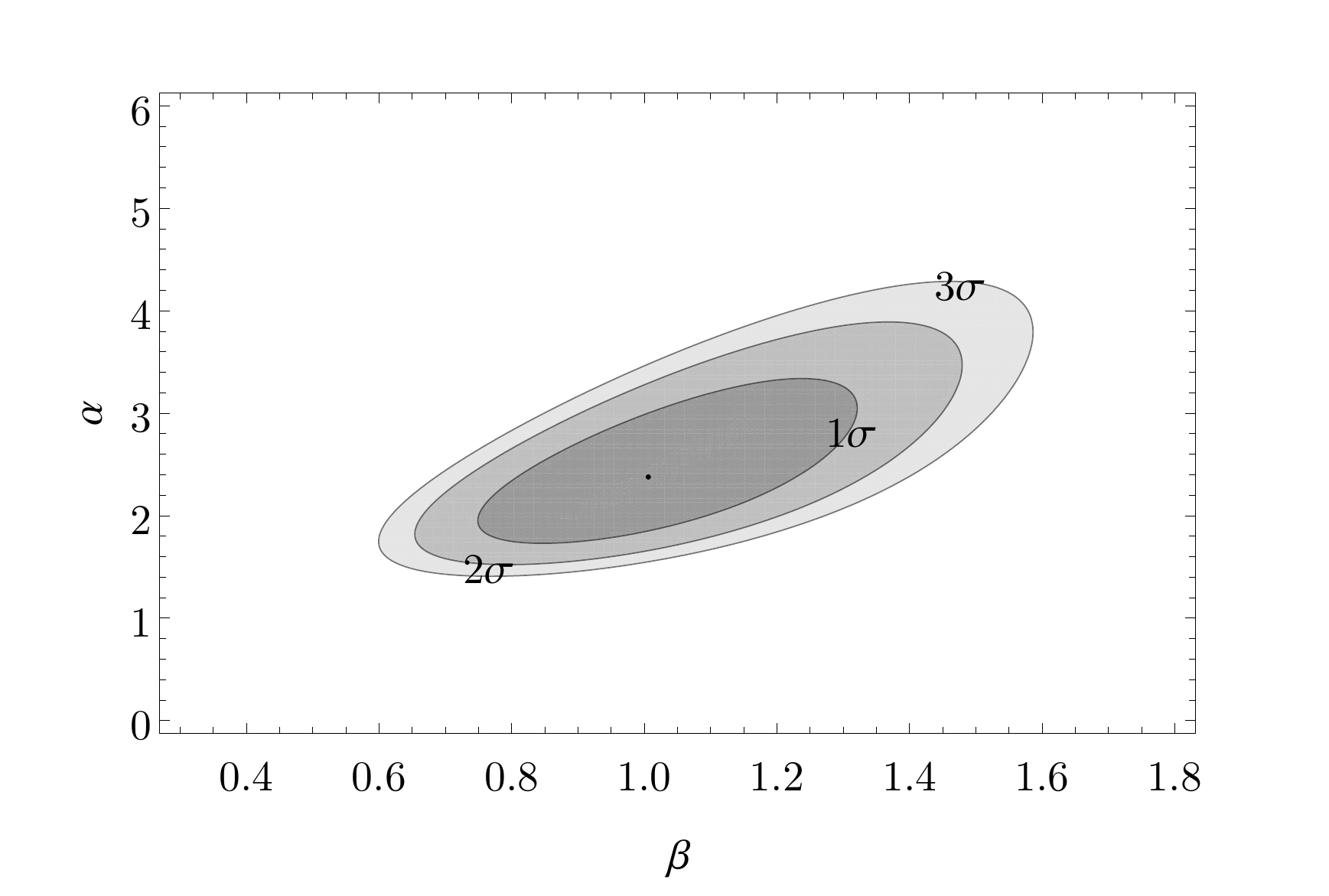}}
\subfloat[ $\Lambda$CDM ]{ \label{Fig:GrafManchaUNBAOb}
\includegraphics[width=0.5\textwidth]{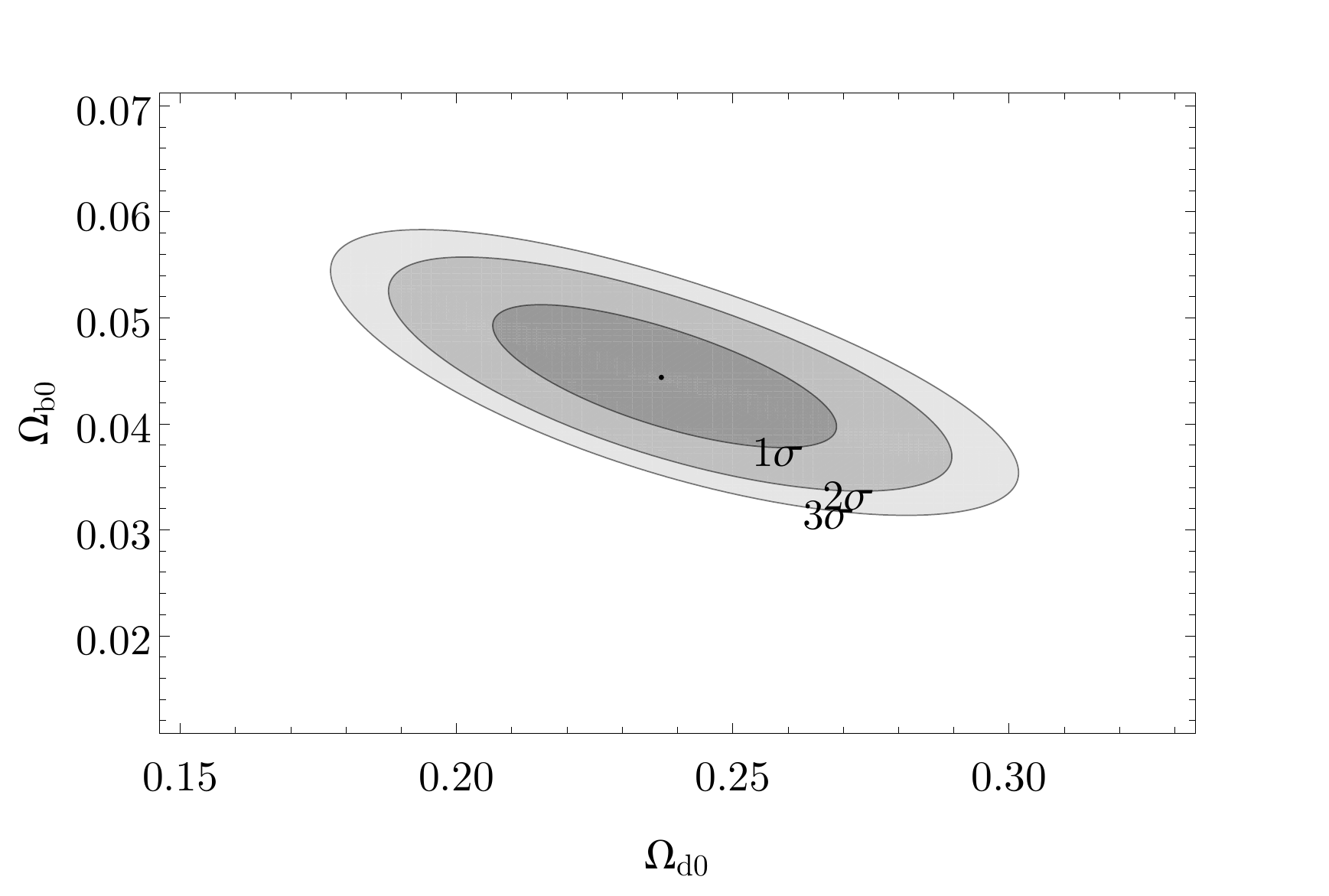}}

\caption{(a) UM: 68.3\%, 95.4\% and 99.7\% confidence regions of the $(\beta , \alpha)$ plane from SNIa plus PN and BAO. (b) $\Lambda$CDM model: 68.3\%, 95.4\% and 99.7\% confidence regions of the $(\Omega_{d0} , \Omega_{b0})$ plane from the combination (SNIa + PN + BAO). Confidence contours include statistical and systematic errors in both cases.}
\label{Fig:NSB_GrafMancha}%
\end{figure*}

\begin{table}[h]

\caption{Parameters of the UM and of the $\Lambda$CDM model obtained through
the fits to the various sets of data. Uncertainties of  $z_{\text{eq}}$ are calculated by the Monte Carlo approach \cite{Watt2012}. All statistical analysis include SNIa and PN data: they are taken as a basis for
comparison with the results coming from the addition of GRB and BAO data.}
\label{Tab:Parametros}

\renewcommand{\arraystretch}{1.3}

\begin{tabular}{c|c|cc|cc|cc|cc}
\hline
\multirow{2}{*}{Set} & \multirow{2}{*}{Parameter} & \multicolumn{2}{c|}{PN+SNIa}  & \multicolumn{2}{c|}{PN+SNIa+GRB} & \multicolumn{2}{c|}{PN+SNIa+BAO}  & \multicolumn{2}{c}{PN+SNIa+GRB+BAO} \\ 
                        &                            & Best-fit  & Single-parameter   & Best-fit  & Single-parameter   & Best-fit  & Single-parameter   & Best-fit  & Single-parameter \\ \hline
                       
\multirow{5}{*}{UM} 

 & $\alpha$ & 2.1 & $2.1_{-1.2}^{+1.6}$ & 2.4 & $2.6_{-1.1}^{+1.5}$  & 2.4 & $2.39_{-0.37}^{+0.43}$ & 2.4 & $2.42_{-0.38}^{+0.44}$ \\  
 & $\beta$ & 0.92 & $0.95_{-0.34}^{+0.45}$ & 0.99 & $1.05_{-0.32}^{+0.45}$ & 1.0 & $1.02_{-0.14}^{+0.15}$ & 0.99 & $1.01_{-0.14}^{+0.15}$ \\ 
 & $\Omega_{b0}$ & 0.041 & $0.0412_{-0.0085}^{+0.0073}$  & 0.041 & $0.0412_{-0.0085}^{+0.0073}$ & 0.041 & $0.0401_{-0.0080}^{+0.0070}$ & 0.041 & $0.0393_{-0.0080}^{+0.0070}$ \\
 & $z_{\text{eq}}$ & 0.45 & $0.377_{-0.066}^{+0.103}$ & 0.41 & $0.366_{-0.050}^{+0.069}$ & 0.42 & $0.409_{-0.046}^{+0.058}$ & 0.41 & $0.404_{-0.049}^{+0.048}$ \\
 & $\chi^2_{\text{red}}$ & 0.97 & - & 0.94 & - & 0.96 & - & 0.94 & - \\ 
\hline\hline

\multirow{5}{*}{$\Lambda$CDM}

 & $\Omega_{d0}$ & 0.24 & $0.237_{-0.021}^{+0.021}$  & 0.24 & $0.244_{-0.020}^{+0.020}$ & 0.24 & $0.237_{-0.020}^{+0.021}$  & 0.24 & $0.243_{-0.020}^{+0.020}$ \\  
 & $\Omega_{b0}$ & 0.041 & $0.0407_{-0.0079}^{+0.0078}$  & 0.041 & $0.0407_{-0.0079}^{+0.0078}$ & 0.044 & $0.0443_{-0.0044}^{+0.0045}$  & 0.044 & $0.0435_{-0.0043}^{+0.0044}$ \\ 
 & $z_{\text{eq}}$ & 0.45 & $0.444_{-0.054}^{+0.058}$ & 0.43 & $0.430_{-0.054}^{+0.050}$  & 0.45 & $0.442_{-0.053}^{+0.055}$ & 0.43 & $0.429_{-0.052}^{+0.049}$ \\ 
 & $\chi^2_{\text{red}}$ & 0.97 & - & 0.94 & - & 0.96 & - & 0.94 & - \\

\hline\hline

\end{tabular}

\renewcommand{\arraystretch}{1.2}

\end{table}

\end{widetext}

\subsection{PN, SNIa, GRB and BAO data sets\label{All data sets.}}

Our final statistical analyzes takes into account all the data sets: primordial nucleosynthesis constraint, type Ia supernovae, gamma-ray bursts and baryon acoustic oscillations. The best-fit parameter are shown in Table \ref{Tab:Parametros}. 




\begin{figure*}[!t]

\subfloat[Unified Model]{ \label{Fig:GrafManchaMUAll}
\includegraphics[width=0.5\textwidth]{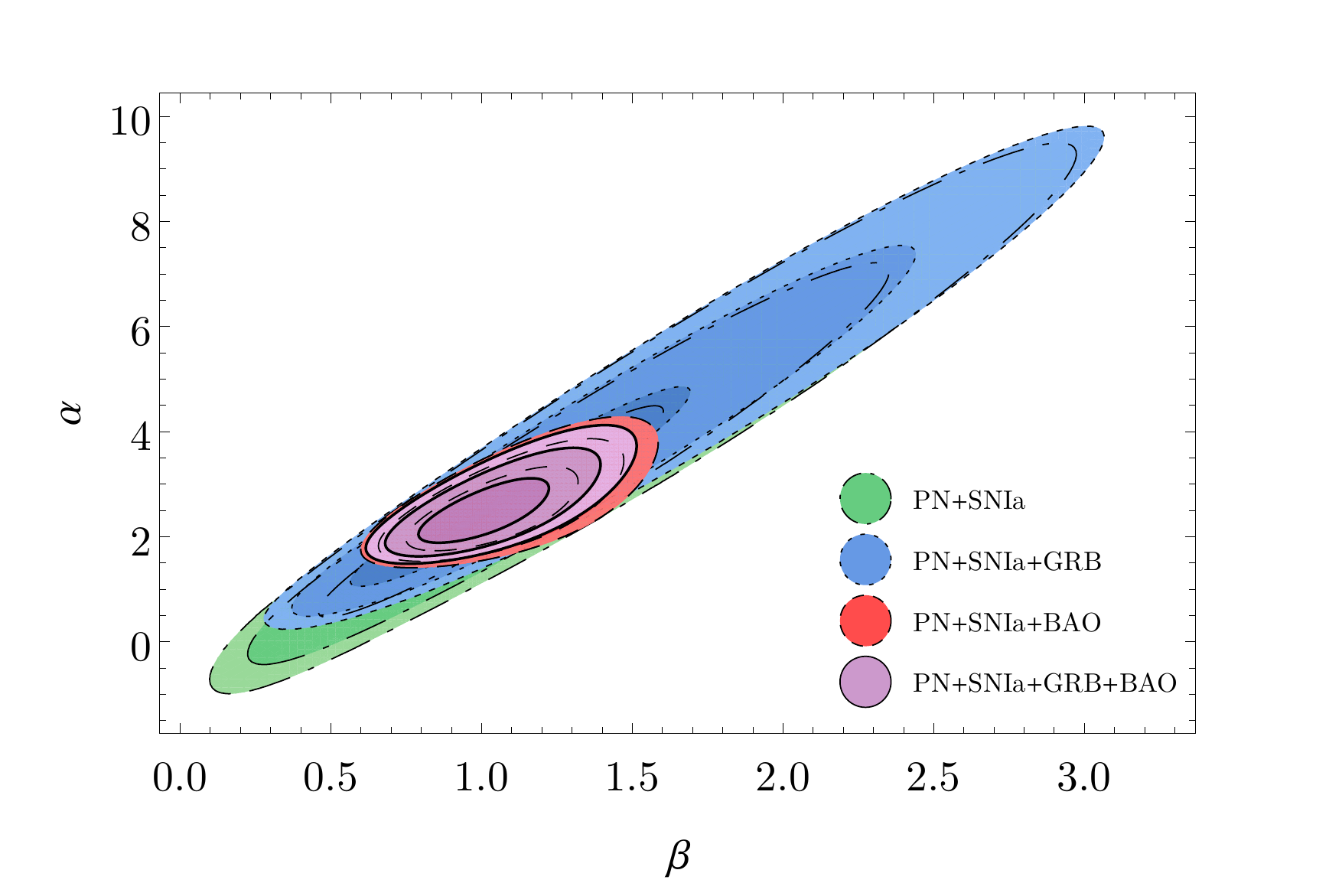}}
\subfloat[ $\Lambda$CDM ]{ \label{Fig:GrafManchaLCDMAll}
\includegraphics[width=0.5\textwidth]{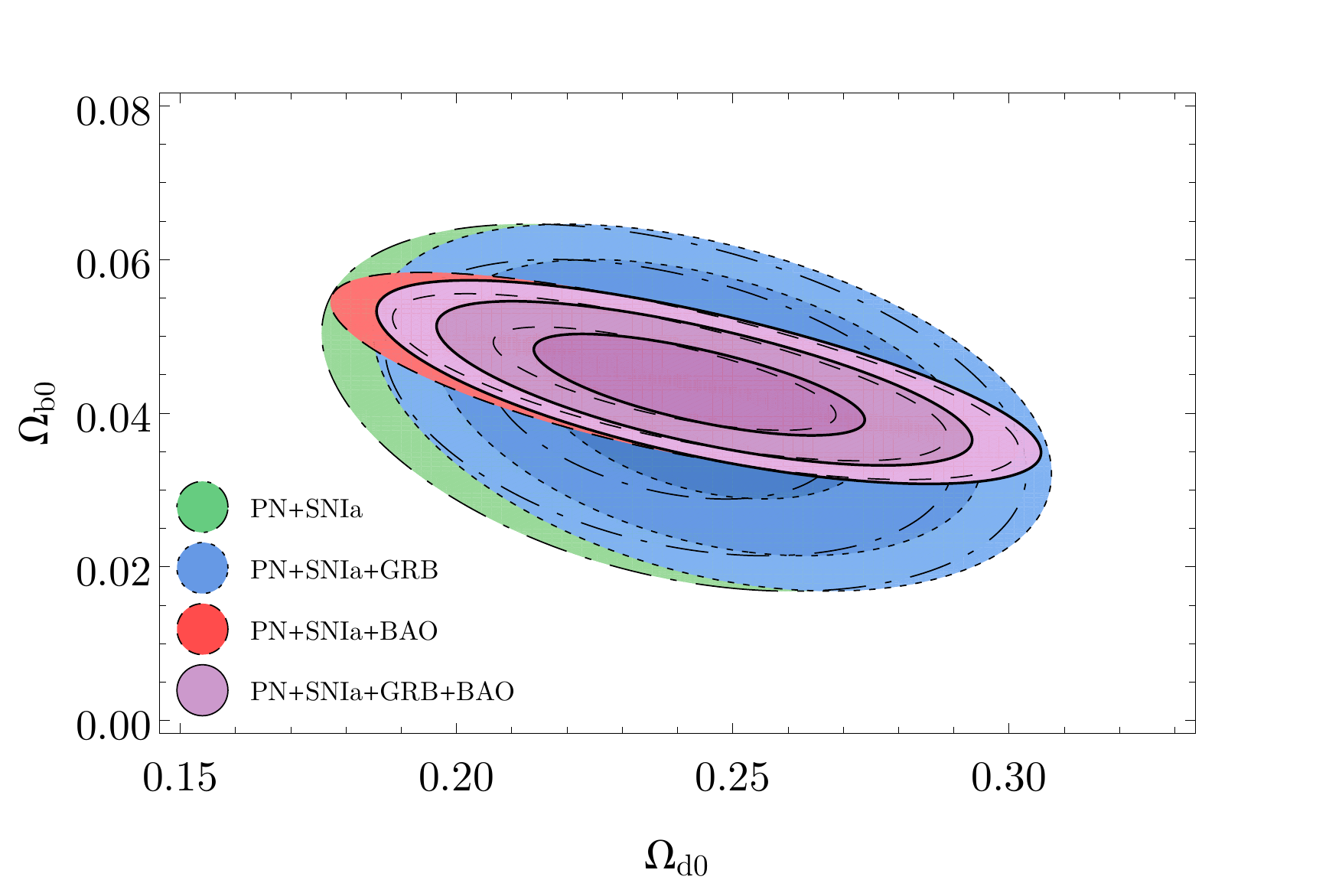}}

\caption{Confidence contours for the planes $(\beta, \alpha)$ of the UM (Fig. \ref{Fig:GrafManchaMUAll}) and $(\Omega_{d0},\Omega_{b0})$ of $\Lambda$CDM model (Fig. \ref{Fig:GrafManchaLCDMAll}) for various sets of data. The double-parameter estimates are little affected by the inclusion of GRB data in the case of both models: the $3\sigma$-confidence region of the set (PN + SNIa + GRB) is slightly smaller than the one for the set (PN + SNIa). Conversely, the confidence regions of the two-parameter estimates are considerably reduced by the inclusion of the BAO peaks in the statistical treatment. }
\label{Fig:GrafMancha}%
\end{figure*}

Fig. \ref{Fig:GrafMancha} shows the confidence regions of the plane of parameters for the UM and $\Lambda$CDM model built with all data sets (SNIa + PN + GRB + BAO). It simultaneously displays the confidence contours of the previous analyses in order to indicate the impact of the different data set in constraining the domain value of the parameters.

\begin{figure*}[t]

\subfloat[\hspace{-1.0cm} ]{ \label{Fig:compWeffDarka}
\includegraphics[width=0.5\textwidth]{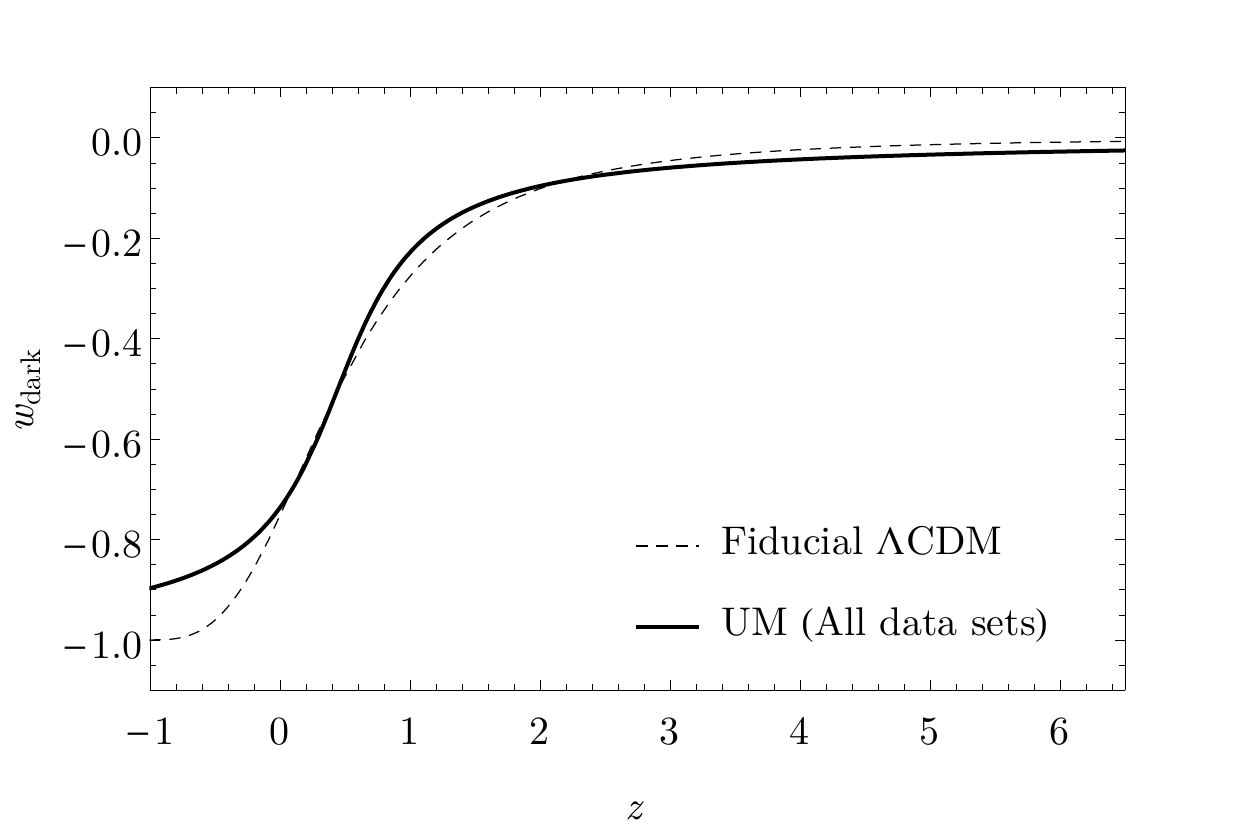}}
\subfloat[\hspace{-1.0cm} ]{ \label{Fig:compWeffDarkb}
\includegraphics[width=0.5\textwidth]{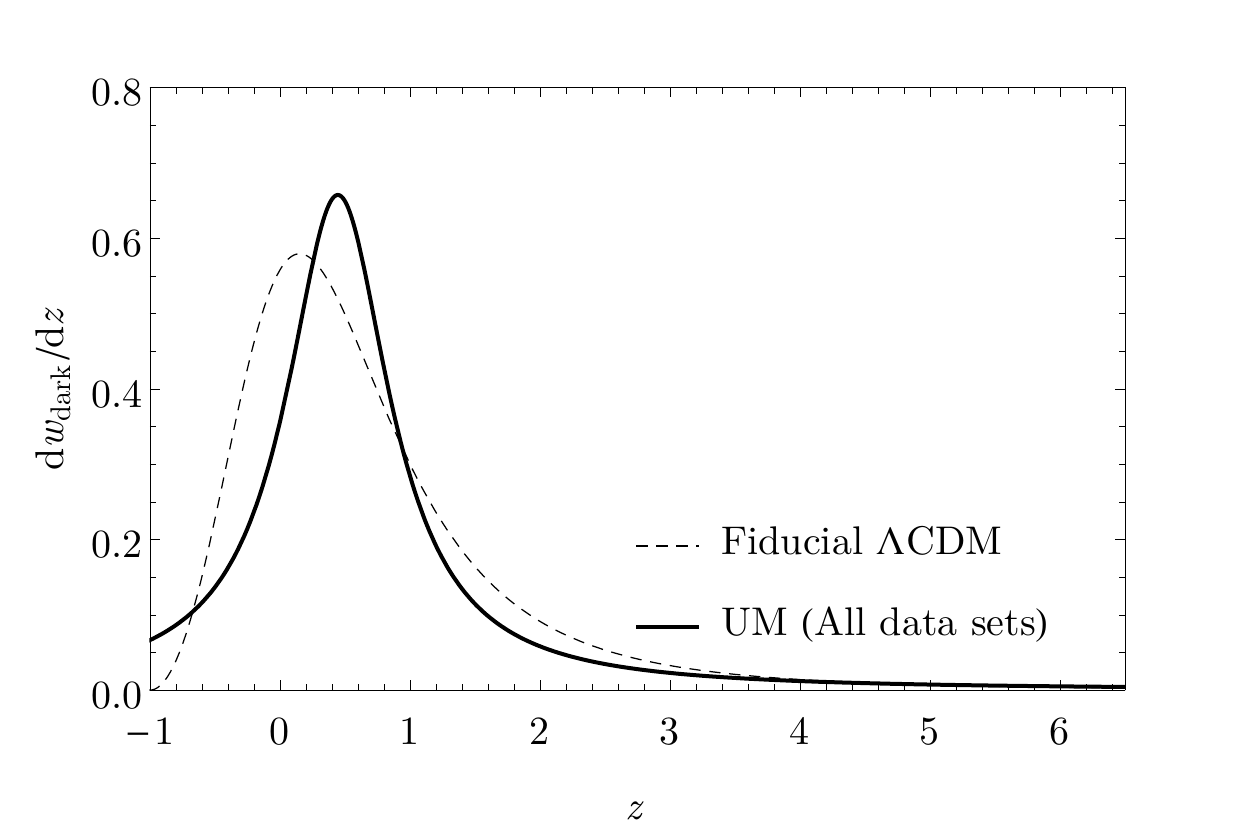}}

\caption{Comparison between fiducial $\Lambda$CDM model ($\Omega_{b0}=0.04$;
$\Omega_{d0}=0.24$) and the unified model with the best-fit values of
parameters $\alpha,$\ $\beta$\ and $\Omega_{b0}$\ obtained with SNIa, GRB, BAO
and PN data sets ($\alpha=2.4$; $\beta=0.99$; $\Omega_{b0}=0.041$).
Plots of (a) $w_{\text{dark}}$\ and (b) of $dw_{\text{dark}}/dz$\ \ with respect
to redshift for both models.}%
\label{Fig:CompWeff}%
\end{figure*}

Assuming that the set (PN + SNIa + GRB + BAO) gives the most realistic values
for the cosmological parameters, we use the best-fit results for $\alpha$\ and
$\beta$ in Eq. (\ref{Eq:weffMU}) in order to obtain $w_{\text{U}}=w_{\text{dark}}$\ of the
dark sector of the universe according to UM. We also obtain
$w_{\text{dark}}$\ of the dark components in the $\Lambda$CDM model,
using\emph{\ }$\Omega_{d0}=0.24$\ e $\Omega_{b0}=0.04$, for comparison. Both
models are characterized by $w_{\text{dark}}(z)$ whose behavior are shown in Fig \ref{Fig:compWeffDarka}. In the distant future, one anticipates $a\rightarrow\infty$,\ which implies $z\rightarrow-1$. The $\Lambda$CDM model gives $w_{\text{dark}}\left(  z=-1\right)
=-1$\ while\ the UM leads to $w_{\text{dark}}\left(  z=-1\right)  =-0.90$.
Notice that there are no big differences between the models in the region of
small redshifts ($0\lesssim z\lesssim0.5$). In addition, the present-day
values ($z=0$) are $w_{\text{dark},0}=-0.75$\ for the $\Lambda$CDM model and
$w_{\text{dark},0}=-0.74$\ for the UM. The functions $w_{\text{dark}}\left(
z\right)  $\ of both models are equal in the region of $z\approx2.5$ and
slightly different elsewhere. Fig. \ref{Fig:compWeffDarkb} shows that the
transition rates $dw_{\text{dark}}/dz$\ for the two models are well
distinguished. The peak of the transition rate for the $\Lambda$CDM model is
$\left(  dw_{\text{dark}}/dz\right)  _{\max}=0.58$\ and occurs at $z_{\max
}=0.15$. For the UM we get $\left(  dw_{\text{dark}}/dz\right)  _{\max}=0.66$\ at the larger redshift of $z_{\max}=0.44$. Both models interchange the quality of being the one with the larger transition rate depending on the value of $z$.

\section{Final comments \label{Sec-FinalComments}}

This work presented a cosmological model unifying dark matter and dark energy
through a parameterization in terms of the function $\arctan$. The three parameters
of the model, $\alpha$, $\beta$ and $\Omega_{b0}$, were estimated admitting
flat spacial curvature and using four observational data set, namely: PN,
SNIa, GRB and BAO. The same combination of data was employed to constraint the
$\Lambda$CDM\ model. This was used as standard with respect to which our model
was compared.

The results were analyzed in two distinct ways: (i) the influence of the
inclusion of GRB and BAO data in the estimates of the parameters of UM
and $\Lambda$CDM model was discussed, and (ii) the direct comparison of UM
and $\Lambda$CDM model was performed. In regard to point (i), it can be
said that the inclusion of GRB data to the basic set (PN plus SNIa) does not
modifies in a decisive way the confidence contours. In fact, there is a small difference
between the curves in Fig. \ref{Fig:GrafMancha} even after increasing the
number of GRB (by including 29 GRB to the set presented in \cite{Wei3}) and
improving the interpolation technique of the calibration procedure. This
indicates that, in spite of been promising as standard candles, GRB events
are still not competitive in comparison to other sets of data such as the one for supernovae. Unlike the GRB data, the inclusion of BAO significantly restricts
the parameter space; this is particularly true for the Unified Model (see Fig.
\ref{Fig:GrafManchaMUAll}). With respect to point (ii), we can say that the UM
and $\Lambda$CDM model exhibit statistically equivalent results for the baryon
density $\Omega_{b0}$ and the redshift $z_{\text{eq}}$. Moreover,
$\chi_{\text{red}}^{2}$ for both models are practically the same. In addition,
the cosmic dynamics of the two models are very similar on the best-fit for all
$z\geq0$ (cf. Fig. \ref{Fig:compWeffDarka}). The most pronounced difference
between UM and $\Lambda$CDM occurs in their evolution toward the future, for
$-1<z<0$. In fact, our parameterization leads to $\lim_{z\rightarrow-1}w_{\text{dark}
}=-0.90$ and not to $\lim_{z\rightarrow-1}w_{\text{dark}}=-1$ as in the
fiducial model. We can not affirm at the current stage of our investigation, if this difference between models is a physical effect due to the unification of the dark components in the UM
or only an artifact of the parameterization for $w$ that we have chosen.

Future perspectives include two important subjects. The first concerns the
dependence of the results on the specific parameterization for $w\left(
z\right)  $\ chosen in our Unified Model. In particular, the $\arctan$
parameterization does not contain the $\Lambda$CDM model, i.e. there is no
combination of the values of $\alpha$\ and $\beta$ leading to the
$w_{\text{dark}}$ of the $\Lambda$CDM\ model. This issue might be overcome by
employing other parameterization such as one based on function $\tanh$. A
second matter of investigation would be a possible UM-$\Lambda$CDM equivalence
in a perturbative level. Indeed, could the statistical equivalence encountered
in our data analysis (performed on the background) show up in a perturbative
approach as well? These two question shall be addressed in further works.

\begin{acknowledgments}
RRC and EMM are grateful to FAPEMIG-Brazil (grant CEX--APQ--04440-10) for
financial support. EMM thanks CAPES-Brazil for financial support. LGM acknowledges FAPERN-Brazil for financial support.
\end{acknowledgments}

\appendix

\section{Akima's interpolation method\label{Akima}}

Akima proposed in \cite{Akima} a new interpolation technique aiming to overcome a difficulty
shared by other interpolation methods, namely: the curve intercepting the data
set does not present a natural evolution, as if it were drawn by hand.
Typically, these other methods violate the continuity of the function or of
its first-order derivative in some region of the domain; even if this flaw does
not occur, the resulting curve presents undesirable oscillations or instabilities.

Ref. \cite{Akima} establishes an interpolation method based in a piecewise
function built with third-degree polynomials. The continuity of the composite
function and its derivative are guaranteed by geometrical arguments. The slope
$t$\ of a given intermediate point among five neighboring points is calculated
by%
\begin{equation}
t=\dfrac{m_{2}\left\vert m_{4}-m_{3}\right\vert +m_{3}\left\vert m_{2}%
-m_{1}\right\vert }{\left\vert m_{4}-m_{3}\right\vert +\left\vert m_{2}%
-m_{1}\right\vert }~, \label{Eq:Ap1}%
\end{equation}
where $m_{i}$\ is the slope of the straight line connecting the $i$-th point
(among the five points of the set) to the $\left(  i+1\right)  $-th
point. For instance, $m_{2}$\ is the angular coefficient of the
straight line connecting the second and third points. The slopes uncertainties
are obtained through the method of propagation of uncertainties after a long but straightforward calculation.

By using Eq. (\ref{Eq:Ap1}), one estimates the slopes for a set with $N$
points ($x_{i}$, $y_{i}$)\ except for the four points at the ends. Then, a
third-degree polynomial is interpolated to the neighboring points respecting
their coordinates and the determined slopes. Notice that by knowing the two
coordinates and the two derivatives associated to a pair of points we are able
to interpolate a third-degree polynomial, which has four degrees of freedom.
However, we can not estimate the rate of change of the two last points at the
ends using (\ref{Eq:Ap1}). These extremal points are interpolated to their
internal neighbors, whose coordinates and slopes are known.

\begin{figure}

\includegraphics[width=0.5\textwidth]{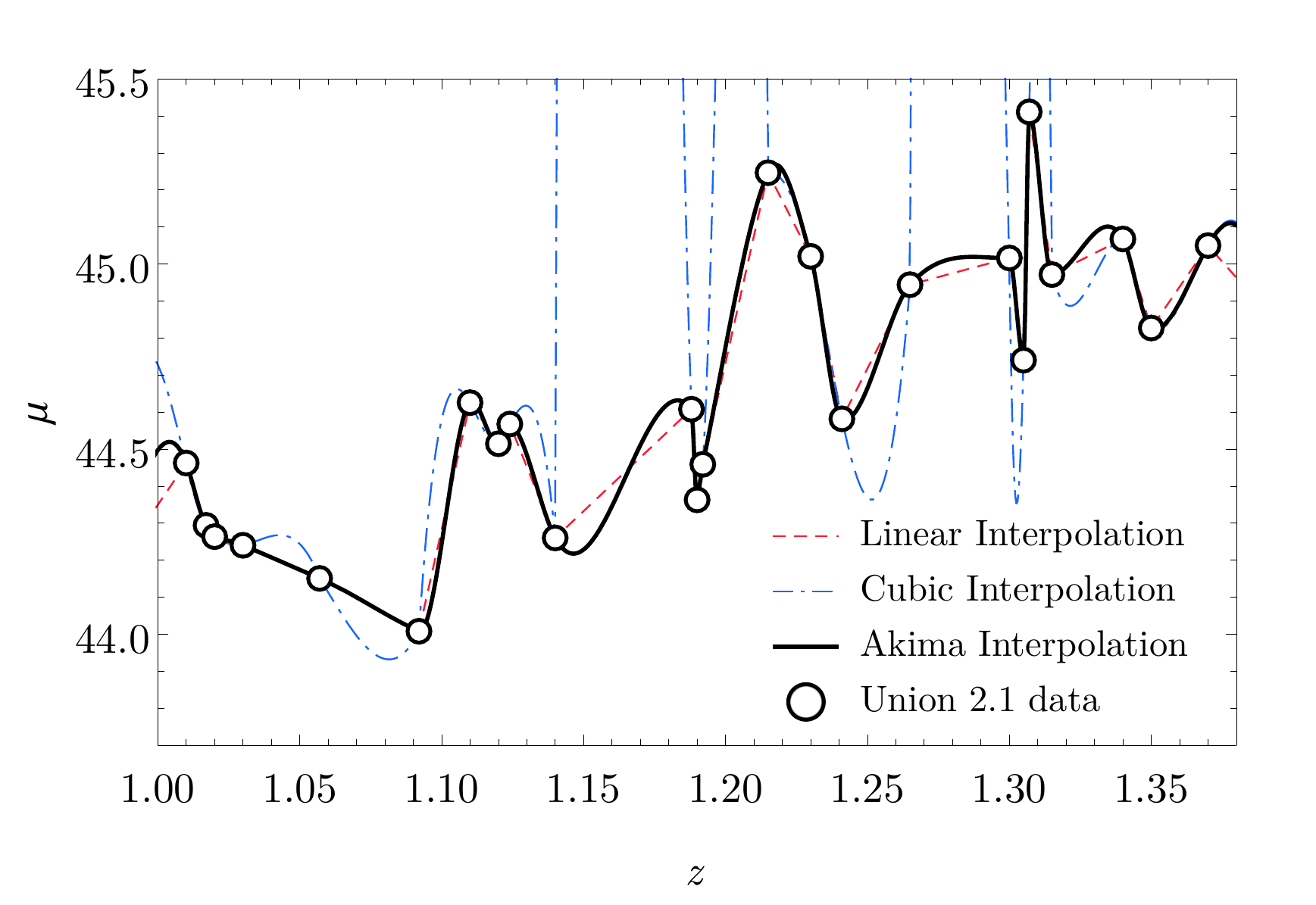}

\caption{Comparison of the linear and cubic interpolation method with Akima's
technique. We use part of the Union2.1 data.}%
\label{Fig:Interpolation}%
\end{figure}

Fig. \ref{Fig:Interpolation} shows part of the interpolation curves built
according to linear, cubic and Akima's interpolation methods. The zoom
includes Union2.1 data from redshift $1$ to $1.4$. The linear interpolation
produces a curve connecting the points in a direct form; but the first-order
derivative of the function describing the curve is not continuous at the points.

On the other hand, cubic interpolation generates a smooth function with a
continuous first-order derivative. However, huge instabilities and oscillations
show up (such as those between point number 10 and point number 11 in the
sample). This makes this method unsuited for the process of calibrating GRB curves.

For this end, Akima's interpolation is the most adequate one because it gives a
smooth and continuous function; this function has continuous derivative; and,
the interpolated function follows the natural tendency of the points and in
between them (i.e. there are no spurious oscillations).

\section{High-redshift GRB distance modulus\label{App:TabelaHighGRB}}

In subsection \ref{subsec:CalibrarGRB}, we have calibrated the $79$%
\ high-redshift GRB compiled in Ref. \cite{Liu2014b}. Here, the results are
presented in Table \ref{GRBTable}.

\bigskip

\begin{longtable*}{C{2.5cm} C{2.5cm} C{2.5cm}}
\caption[GRB's de alto redshift]{Distance modulus data for high-redshift supernovae.} \label{GRBTable} \\
\hline \hline \multicolumn{1}{c}{\textit{z }} & \multicolumn{1}{c}{\textbf{$\mu$}} & \multicolumn{1}{c}{\textbf{$\sigma_{\mu}$ }} \\ \hline
\endfirsthead
\multicolumn{3}{c}%
{{ \tablename\ \thetable{} -- continued from previous page}} \\
\hline \textit{z } &\textbf{$\mu$} &\textbf{$\sigma_{\mu}$ }\\ \hline
\endhead
\hline \multicolumn{3}{r}{{Continued on next page.}} \\
\endfoot
\hline \hline
\endlastfoot
\rowcolor{Cor1}
1.44 & 43.68 & 1.02 \\
1.44 & 44.18 & 1.08 \\
\rowcolor{Cor1}
1.46 & 44.41 & 1.00 \\
1.48 & 43.97 & 1.00 \\
\rowcolor{Cor1}
1.49 & 45.43 & 1.12 \\
1.52 & 43.26 & 1.04 \\
\rowcolor{Cor1}
1.55 & 44.48 & 1.04 \\
1.55 & 46.33 & 1.05 \\
\rowcolor{Cor1}
1.56 & 43.15 & 1.77 \\
1.60 & 44.60 & 1.13 \\
\rowcolor{Cor1}
1.60 & 47.03 & 1.04 \\
1.61 & 47.38 & 1.13 \\
\rowcolor{Cor1}
1.62 & 44.77 & 1.02 \\
1.64 & 45.31 & 1.01 \\
\rowcolor{Cor1}
1.71 & 47.45 & 1.66 \\
1.73 & 43.64 & 1.05 \\
\rowcolor{Cor1}
1.80 & 45.86 & 1.04 \\
1.82 & 45.25 & 1.00 \\
\rowcolor{Cor1}
1.90 & 46.25 & 1.19 \\
1.95 & 46.95 & 1.16 \\
\rowcolor{Cor1}
1.97 & 45.07 & 1.06 \\
1.98 & 44.94 & 1.08 \\
\rowcolor{Cor1}
2.07 & 44.35 & 1.03 \\
2.10 & 47.16 & 1.37 \\
\rowcolor{Cor1}
2.11 & 47.42 & 1.01 \\
2.11 & 44.64 & 1.00 \\
\rowcolor{Cor1}
2.14 & 45.19 & 1.03 \\
2.15 & 47.83 & 1.15 \\
\rowcolor{Cor1}
2.20 & 46.81 & 1.17 \\
2.20 & 47.26 & 1.01 \\
\rowcolor{Cor1}
2.22 & 45.32 & 1.18 \\
2.30 & 45.91 & 1.22 \\
\rowcolor{Cor1}
2.30 & 46.59 & 1.31 \\
2.35 & 47.27 & 1.22 \\
\rowcolor{Cor1}
2.35 & 46.74 & 1.36 \\
2.43 & 46.82 & 1.06 \\
\rowcolor{Cor1}
2.43 & 47.35 & 1.18 \\
2.45 & 47.86 & 1.21 \\
\rowcolor{Cor1}
2.51 & 46.92 & 1.05 \\
2.58 & 45.55 & 1.03 \\
\rowcolor{Cor1}
2.59 & 46.62 & 1.04 \\
2.61 & 46.32 & 1.07 \\
\rowcolor{Cor1}
2.65 & 46.02 & 1.07 \\
2.69 & 46.44 & 1.12 \\
\rowcolor{Cor1}
2.71 & 45.27 & 1.33 \\
2.75 & 45.85 & 1.13 \\
\rowcolor{Cor1}
2.77 & 45.99 & 1.00 \\
2.82 & 47.05 & 1.01 \\
\rowcolor{Cor1}
2.90 & 45.73 & 1.11 \\
3.00 & 46.63 & 1.18 \\
\rowcolor{Cor1}
3.04 & 46.55 & 1.03 \\
3.04 & 45.38 & 1.25 \\
\rowcolor{Cor1}
3.08 & 47.55 & 1.20 \\
3.20 & 46.23 & 1.18 \\
\rowcolor{Cor1}
3.21 & 45.96 & 1.19 \\
3.34 & 47.49 & 1.06 \\
\rowcolor{Cor1}
3.35 & 48.09 & 1.03 \\
3.36 & 45.82 & 1.04 \\
\rowcolor{Cor1}
3.37 & 47.81 & 1.32 \\
3.42 & 47.45 & 1.07 \\
\rowcolor{Cor1}
3.43 & 47.18 & 1.02 \\
3.53 & 47.15 & 1.03 \\
\rowcolor{Cor1}
3.57 & 46.35 & 1.06 \\
3.69 & 45.74 & 1.07 \\
\rowcolor{Cor1}
3.78 & 49.24 & 1.41 \\
3.91 & 46.71 & 1.18 \\
\rowcolor{Cor1}
4.05 & 48.52 & 1.04 \\
4.11 & 47.39 & 1.27 \\
\rowcolor{Cor1}
4.27 & 48.13 & 1.23 \\
4.35 & 47.57 & 1.10 \\
\rowcolor{Cor1}
4.41 & 48.47 & 1.07 \\
4.50 & 46.55 & 1.29 \\
\rowcolor{Cor1}
4.90 & 47.43 & 1.25 \\
5.11 & 48.67 & 1.07 \\
\rowcolor{Cor1}
5.30 & 47.89 & 1.05 \\
5.60 & 48.45 & 1.02 \\
\rowcolor{Cor1}
6.29 & 50.02 & 1.20 \\
6.70 & 50.27 & 1.39 \\
\rowcolor{Cor1}
8.10 & 49.75 & 1.29 \\
\end{longtable*}

\bigskip

\bibliography{References}

\end{document}